\documentclass[preprint2]{aastex}
\usepackage{natbib}
\usepackage{amsmath}
\usepackage{color}
\usepackage{longtable}
\usepackage{enumerate}
\usepackage{mathtools}

\shorttitle{D-organics in Orion-KL}
\shortauthors{Favre et al.}

\begin{document}

\title{The distribution of deuterated formaldehyde within Orion--KL}

\author{C\'ecile Favre, Edwin~A. Bergin, Justin~L. Neill,}
\affil{Department of Astronomy, University of Michigan, 500 Church St., 
\email{cfavre@umich.edu}
    Ann Arbor, MI 48109, USA}
   
\and

\author{Nathan~R. Crockett}
\affil{California Institute of Technology, Division of Geological and Planetary Sciences, MS 150-21, Pasadena, CA 91125, USA}
        
\and

\author{Qizhou Zhang}
\affil{Harvard-Smithsonian Center for Astrophysics, 60 Garden Street, Cambridge MA 02138, USA}
 
\and

\author{Dariusz~C. Lis}
\affil{LERMA, Observatoire de Paris, PSL Research University, CNRS, Sorbonne Universit\'es, UPMC Univ. Paris 06, F-75014, Paris, France}
\affil{Cahill Center for Astronomy and Astrophysics 301--17, California Institute of Technology, Pasadena, CA 91125, USA}

%
%
\begin{abstract}
We report the first high angular resolution imaging (3.4\arcsec $\times$ 3.0\arcsec) of deuterated formaldehyde (HDCO) toward Orion--KL, carried out with the Submillimeter Array (SMA). We find that the spatial distribution of the formaldehyde emission systematically differs from that of methanol: while methanol is found towards the inner part of  the region, HDCO is found in colder gas that wraps around the methanol emission on four sides. The HDCO/H$_2$CO ratios are determined to be 0.003--0.009 within the region, up to an order of magnitude higher than the D/H measured for methanol. These findings  strengthen the previously suggested hypothesis  that there are differences in the chemical pathways leading to HDCO (via deuterated gas phase chemistry) and deuterated methanol (through conversion of formaldehyde into methanol on the surface of icy grain mantles). 
\end{abstract}


%
\section{Introduction}
\label{sec:introduction}

The Orion--KL nebula, that lies at a distance of 414$\pm$7~pc \citep{Menten:2007} is a focal source for studies of the physics and chemistry of high--mass star forming regions. Its rich molecular diversity and complex physical structure have been revealed by a number of spectral line surveys and interferometric studies of the region \citep[e.g.][]{Blake:1987,Comito:2005,Schilke:1997,Schilke:2001,Beuther:2005,Beuther:2006,Persson:2007,Guelin:2008,Friedel:2008,Tercero:2010,Tercero:2011,Favre:2011,Favre:2011a,Favre:2014a,Brouillet:2013,Peng:2013,Friedel:2008,Friedel:2012,Widicus-Weaver:2012,Esplugues:2013}. Recently, a broadband spectral line survey of Orion--KL carried out with the Herschel/HIFI spectrometer \citep{Bergin:2010,Crockett:2010,Crockett:2014} has been obtained as part of the HEXOS key program\footnote{www.hexos.org}. The unprecedented 1.2~THz wide frequency coverage of this survey has allowed for strong constraints on physical parameters and molecular abundances due to the detection of many transitions for each molecule, with a wide range of excitation conditions, revealing the presence of thermal gradients \citep[e.g.][]{Wang:2011}. Among the molecular components associated with Orion--KL lie the so-called Hot Core and Compact Ridge. Although the precise nature of these regions remains enigmatic, evidence has been recently provided that they may have originated from the interaction of the remnants of a recent explosion with ice grains mantles \citep[$\le$720 years ago, e.g.][]{Bally:2011,Gomez:2005,Gomez:2008,Rodriguez:2005,Nissen:2012}, leading to the evaporation and the heating of organic material \citep{Favre:2011,Zapata:2011}.

Deuterium fractionation is intrinsically a low-temperature process. Therefore, deuterated molecules in warm (T$\ge$100K) region, such as Orion-KL, offer a window into its physical conditions at the earlier, colder stage (T$\sim$10--30~K) when this material likely formed \citep{Blake:1987,Charnley:1997,Pagani:1992}.
Numerous deuterated species have been observed toward Orion--KL:  deuterated water \citep{Turner:1975,Jacq:1990,Neill:2013}, deuterated ammonia \citep{Rodriguez-Kuiper:1978,Walmsley:1987}, deuterated formaldehyde \citep{Loren:1985,Turner:1990} and deuterated methanol \citep{Mauersberger:1988,Jacq:1993,Peng:2012}. Recently, \citet{Neill:2013a} used the HIFI survey, and tens of transitions, to set strong constraints in the D/H ratio of organics associated with Orion--KL.
Their finding strongly suggests that D/H fractionation is inhomogenuous in this source.

In this study, we extend the HIFI study of deuterated species performed by \citet{Neill:2013a,Neill:2013}; especially since the Herschel observations are spatially unresolved, so even with well determined column densities within the beam, there can be large deviations in the deuterium fractionation as a function of spatial position. In Section~\ref{sec:observations}, we present the SMA observations to complement the HIFI analysis. In particular, we focus on formaldehyde (H$_2$CO and HDCO). The D/H ratios of this molecule is key to obtaining a full picture of the oxygen chemistry in this region, as formaldehyde is believed to be an intermediate in the grain--surface synthesis of methanol through sequential reactions of H or D atoms with CO \citep[see][]{Watanabe:2002,Cazaux:2011,Taquet:2012a}. Alternatively, gas phase chemistry might independently lead to the formation of formaldehyde via reactions involving CH$_{2}$D$^{+}$ products under warmer conditions \citep[typically T$>$ 50K, see][]{Wootten:1987,Loren:1985}. Incidentally, the tentative detection of the CH$_{2}$D$^{+}$ ion toward Orion-KL by \citet{Roueff:2013} suggests that gas phase chemistry might be efficient for the HDCO formation. In this analysis we present the first high angular resolution image of the spatial distribution of (deuterated) formaldehyde in Orion--KL and discuss the different chemical pathways that can be involved in its formation. More specifically, results and analysis are presented in Section~\ref{sec:results} and discussed in Section~\ref{sec:discussion}, with conclusions set out in Section~\ref{sec:conc}.

%
\section{Observations and data reduction}
\label{sec:observations}

Observations of Orion-KL were carried out with the SMA\footnote{The Submillimeter Array is a joint project between the Smithsonian Astrophysical Observatory and the Academia Sinica Institute of Astronomy and Astrophysics, and is funded by the Smithsonian Institution and the Academia Sinica.}  \citep{Ho:2004} in its compact configuration on December 12, 2013 for about 6.75h. The phase-tracking centre was $\alpha_{J2000}$ = 05$^{h}$35$^{m}$14$\fs$20, $\delta_{J2000}$ = -05$\degr$22$\arcmin$33$\farcs$00. The new broader-bandwidth SMA capability of the 230~GHz band was used to cover the following frequency ranges: 192.6--194.6~GHz and 197.6--199.6~GHz in LSB band and,  207.6--209.6~GHz and  212.6--214.6~GHz in the USB band. An enormous advantage of this broadband capability is the possibility to observe transitions of both HDCO and H$_2$$^{13}$CO in a single tuning (see Table~\ref{tab1}). The correlator was configured with a uniform spectral resolution over 4~GHz bandwidth in each sideband. Each 4~GHz bandwidth is divided into 48 `chunks', each is further divided into 128 channels with a channel width of 0.812~MHz.
The weather conditions were good and stable with an atmospheric opacity of about 0.14 at 225~GHz.
The SMA data were calibrated using the MIR/IDL package\footnote{https://www.cfa.harvard.edu/$\sim$cqi/mircook.html} \citep{Qi:2007}. The nearby quasars 0423$-$013 and 0510$+$180 were used as  complex gain (phase and amplitude) calibrators.

\begin{figure*}
\centering
\includegraphics[angle=270,width=9.cm]{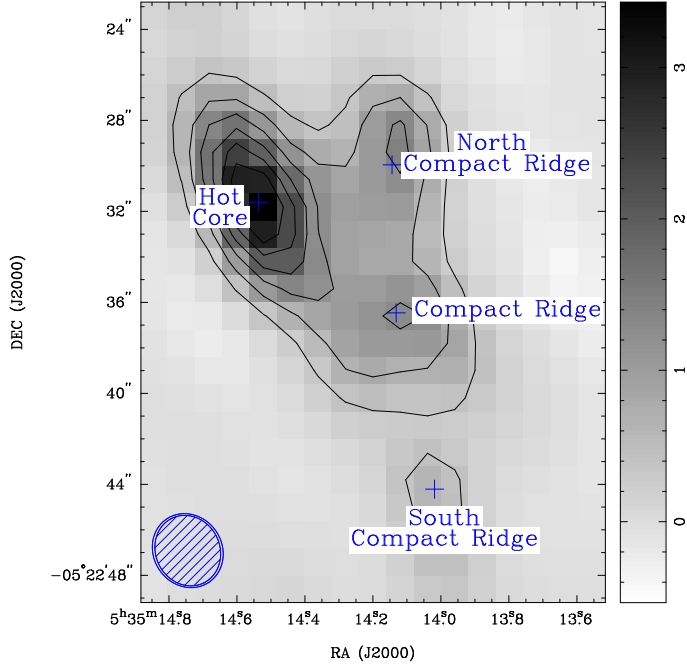}
\caption{Continuum emission maps obtained at 213~GHz (grey scale, USB band) and at 198~GHz (black, LSB band) toward Orion--KL as measured with the SMA. The first contour and the level step are at 3$\sigma$ (where 1$\sigma$=0.132~Jy~beam$^{-1}$). The synthesized beams are 3.2\arcsec $\times$ 2.8\arcsec (P.A. = 26.7$\degr$) and 3.5\arcsec $\times$ 3.01\arcsec (P.A. = 32.7$\degr$) at 207~GHz and 198~GHz, respectively. Blue crosses mark the positions of the four main continuum sources derived from Gaussian fits in the $(u,v)$ plane (198~GHz dataset, LSB band). These sources are  the Hot core ($\alpha_{J2000}$ = 05$^{h}$35$^{m}$14$\fs$537, $\delta_{J2000}$~=~-05$\degr$22$\arcmin$31$\farcs$600), the Compact Ridge ($\alpha_{J2000}$ = 05$^{h}$35$^{m}$14$\fs$132, $\delta_{J2000}$ = -05$\degr$22$\arcmin$36$\farcs$462), the North Compact Ridge ($\alpha_{J2000}$ = 05$^{h}$35$^{m}$14$\fs$144, $\delta_{J2000}$ = -05$\degr$22$\arcmin$29$\farcs$295)  and  the South Compact Ridge ($\alpha_{J2000}$ = 05$^{h}$35$^{m}$14$\fs$019, $\delta_{J2000}$ = -05$\degr$22$\arcmin$44$\farcs$212).}\label{fg1}
\end{figure*}

Continuum subtraction and data imaging were performed using the MIRIAD software package. The final continuum emission maps obtained at 198~GHz and 213~GHz are shown in Figure~\ref{fg1}. They were restored using a robust weighting of 0.0, resulting in a synthesized beam size of 3.5\arcsec $\times$ 3.01\arcsec (P.A. = 32.7$\degr$) at 198~GHz and of 3.2\arcsec $\times$ 2.8\arcsec (P.A. = 26.7$\degr$) at 213~GHz. 
Positions of the four main continuum sources derived from Gaussian fits in the  $(u,v)$ LSB dataset image plane are given in Figure~\ref{fg1}. Positions of these continuum sources are commensurate between the LSB and USB data sets and, in agreement, within the synthesized beam size, with previous PdBI observations \citep{Favre:2011}. 

In this analysis, we only focus on 2 emission lines of HDCO and 2 of para--H$_2$$^{13}$CO (see spectroscopic parameters listed in Table~\ref{tab1}). The  spectral resolution is 1.14~km~s$^{-1}$ for H$_2^{13}$CO at 212/213~GHz and 1.26~km~s$^{-1}$ for the HDCO lines at 193~GHz. The resulting synthesized beam sizes ($\sim$$3\arcsec$$\times$$3\arcsec$, P.A. = 30 $\degr$) are given in Table~\ref{tab1} at the different frequencies.

%
\section{Results and analysis}
\label{sec:results}

\subsection{Emission maps and velocity structure}
\label{sec:map}
The HDCO and H$_2$$^{13}$CO  emission maps integrated over the line profile are shown in  Figure~\ref{fg2}. Four main molecular emission peaks, labeled dF1 to dF4, are associated with the HDCO emission. More specifically, they are located toward the left of Hot Core component (dF1), the north of the Compact Ridge (dF2) and in the quiescent ridge (dF3 and dF4) located south of the Compact Ridge. Their coordinates are given in Table~\ref{tab1}. The regions dF1, dF2 and dF3 are also associated with the H$_2$$^{13}$CO emission but not the dF4 one. Incidentally, it is important to note that the integrated emission map of the H$_2$$^{13}$CO transition at 213293.57~MHz is strongly contaminated by the emission from an unidentified species toward the Hot Core region. This is shown in Fig.~\ref{fg3}, which displays the integrated emission map along with the spectrum of this H$_2$$^{13}$CO line; the latter being taken at the position of the contaminated emission peak.

Spectra of the formaldehyde transitions observed towards the regions dF1 to dF4 are shown in Figure~\ref{fg4} and the observed line parameters are summarized in Table~\ref{tab1}.
The bulk of the formaldehyde emission towards those peaks appears to lie between $v_{LSR}$=7.5~km~s$^{-1}$ and $v_{LSR}$=9.8~km~s$^{-1}$. This result is consistent with the HDCO LSR velocity derived by \citet{Loren:1985} toward the quiescent ridge ($\sim$8.5~km~s$^{-1}$, single-dish Millimeter Wave Observatory observations) and by \citet{Neill:2013a} toward the region ($\sim$8 and 10.4~km~s$^{-1}$, Herschel/HIFI observations). It is important to note that \citet{Neill:2013a} have associated the HDCO emission with two velocity components in Compact Ridge. This difference results from the fact that \textit{i)} some complex molecules, such as methyl formate or dimethyl ether, are seen with two distinct velocity components emitting at $\sim$7.5 and 9.2~km~s$^{-1}$ in the Compact Ridge \citep[e.g.][]{Favre:2011,Brouillet:2013} and that, \textit{ii)} the Herschel observations are spatially unresolved. We infer that the two HDCO velocity components seen by \citet{Neill:2013a}, actually correspond to our HDCO emission observed toward dF3 and dF1, respectively.

\subsection{Excitation temperature}
\label{sec:tempex}
Assuming optically thin emission, we use the integrated line intensity measurements of the observed H$_2$$^{13}$CO transitions to derive the formaldehyde excitation temperature (T$_{ex}$) within Orion--KL following Eq.~\ref{eq1}:
\begin{equation}
\label{eq1}
\rm
\textnormal{T}_{ex} = -\frac{h\nu}{k_{B}ln\left(\frac{g_{l}A_{l}\nu^{2}_{u}W_{u}}{g_{u}A_{u}\nu^{2}_{l}W_{l}}\right)} = -\frac{E_{u} - E_{l}}{ln\left(\frac{g_{l}A_{l}\nu^{2}_{u}W_{u}}{g_{u}A_{u}\nu^{2}_{l}W_{l}}\right)},
\end{equation}
%
where E, W, A, g and $\nu$ refer to the state energies (K), the integrated line intensities (K~km~s$^{-1}$), the Einstein A coefficients (s$^{-1}$), the statistical weights and the line frequencies (Hz) of the 3$_{2, 1}$--2$_{2, 0 }$ (u) and 3$_{0, 3}$--2$_{0, 2}$ (l) transitions. It is important to note that Equation~\ref{eq1} cannot be directly applied to HDCO since both of the detected transitions have the same upper energy level (see Table~\ref{tab1}). Thus, the below analysis hinges upon the assumption that T$_{ex}$ is the same for both HDCO and H$_2$$^{13}$CO.

Figure~\ref{fg5}a presents the excitation temperature map for H$_2$$^{13}$CO within Orion-KL. The formaldehyde excitation temperature derived toward the dF2, dF3 and dF4 peaks is low (52--57~K, see Table~\ref{tab1}) contrary to dF1, for which it is higher ($\sim$90~K). Our derived temperatures in the quiescent ridge (i.e regions dF3 and dF4) are commensurate within the uncertainties, with the one derived by \citet[][60$\pm$30~K]{Bergin:1994} using CH$_3$C$_2$H, a known tracer of ambient ridge gas.
Regarding the dF2 emission peak, our derived temperature is in agreement with one of the KL--W methanol clump that lies in its vicinity \citep[i.e $\sim$2$\arcsec$ east, see][]{Peng:2012}. As for the dF1 emission peak, there is a slight discrepancy with the temperature derived by \citet{Peng:2012}. In particular, these authors report a temperature that is lower than the typical temperature derived in the vicinity of this region \citep[$\sim$100--300~K, e.g.][]{Favre:2011,Goddi:2011a,Friedel:2008}. However, looking at their Figure B.4., a 130~K temperature seems to be consistent with their data, suggesting that their error bars might have been underestimated. Overall  the temperatures derived here are lower than the typical temperatures derived for organic species towards the Hot Core and Compact Ridge components associated with the inner part of Orion--KL \citep[e.g.][]{Crockett:2014}.

\subsection{Column density}
\label{sec:coldens}

Assuming that \textit{i)} LTE is reached, which implies that the excitation temperature is equal to the rotational temperature and, \textit{ii)} the rotational temperature is the same for both HDCO and H$_{2}$$^{13}$CO, we derive the total column density, N,  by using the following equation \citep{Goldsmith:1999}:
\begin{equation}
\label{eq2}
 \rm
ln\left(\frac{8\pi k_{B}\nu^{2}W}{g_{u}hc^{3}A_{ul}}\right)= ln \left(\frac{N}{Q}\right) - \frac{E_{u}}{T} ,\\
\end{equation}
where,  \[
\rm
    \begin{array}{lp{0.8\linewidth}}
    W & is the integrated line intensity (K~cm~$s^{-1}$),  \\
    \nu & the line frequency (Hz), \\
   A_{ul} & the Einstein coefficient (s$^{-1}$),  \\
   g_{u} & the statistical weight, \\
    N & the total column density (cm$^{-2}$), \\
    Q & the partition function, \\
    E_{up} & the upper state energy (K), \\
    h &the Planck constant (erg~s), \\
  k_{B} & the Boltzmann constant (erg/K), \\
    c & the speed of light (cm~$s^{-1}$), and  \\
    T & the excitation temperature derived above (in K, Sect.~3.2). \\
    \end{array}
   \]
   
The column densities derived toward the regions dF1 to dF4 for both HDCO and para--H$_{2}$$^{13}$CO are listed in Table~\ref{tab1}.
In addition, Table~\ref{tab2} gives the total column densities for HDCO, H$_{2}$$^{13}$CO and H$_{2}$CO toward the dF1 to dF4 peaks. The total column densities have been calculated as follow: \textit{i)} the column densities derived for the two HDCO transitions, that are listed in Table~\ref{tab1}, have been averaged. The same applies to the two para--H$_{2}$$^{13}$CO transitions. \textit{ii)} Regarding H$_{2}$$^{13}$CO and HDCO,  we assume an ortho:para ratio of 3:1, that is commensurate with measurements performed towards Orion--KL by \citet{Kahane:1984} and \citet{Crockett:2014}, and a $^{12}$C/$^{13}$C isotopic ratio of 70 \citep[see,][and reference therein]{Favre:2014a}, that is in agreement within the error bars with previous studies \citep[e.g.][]{Persson:2007,Stahl:2008}.

\subsection{D/H ratios for formaldehyde} 

Using the total column densities for HDCO and for H$_{2}$$^{13}$CO (see Section~\ref{sec:coldens} and Table~\ref{tab2}), we derive the D/H ratio for the regions dF1 to dF4. The resulting  D/H ratios are given in Table~\ref{tab2}.
Figure~\ref{fg5}b shows the distribution map of the D/H ratio for formaldehyde in Orion--KL. The HDCO/H$_{2}$CO ratio is found to be 0.009$\pm$0.002 toward dF4, 0.003$\pm$0.001 toward dF1 and 0.004$\pm$0.001 and 0.005$\pm$0.002 toward dF2 and dF3, respectively. Our results are commensurate within the uncertainties with previous D/H ratio measurements in the region \citep[e.g.][]{Loren:1985,Neill:2013a} and suggest a possible gradient in the HDCO/H$_{2}$$^{13}$CO ratio at a 3$\sigma$ level.

%
\section{Discussion}
\label{sec:discussion}
%
%
\subsection{Anticorrelation between formaldehyde and methanol emission}
\label{sec:anti}

\begin{figure*}
\includegraphics[angle=270,width=8cm]{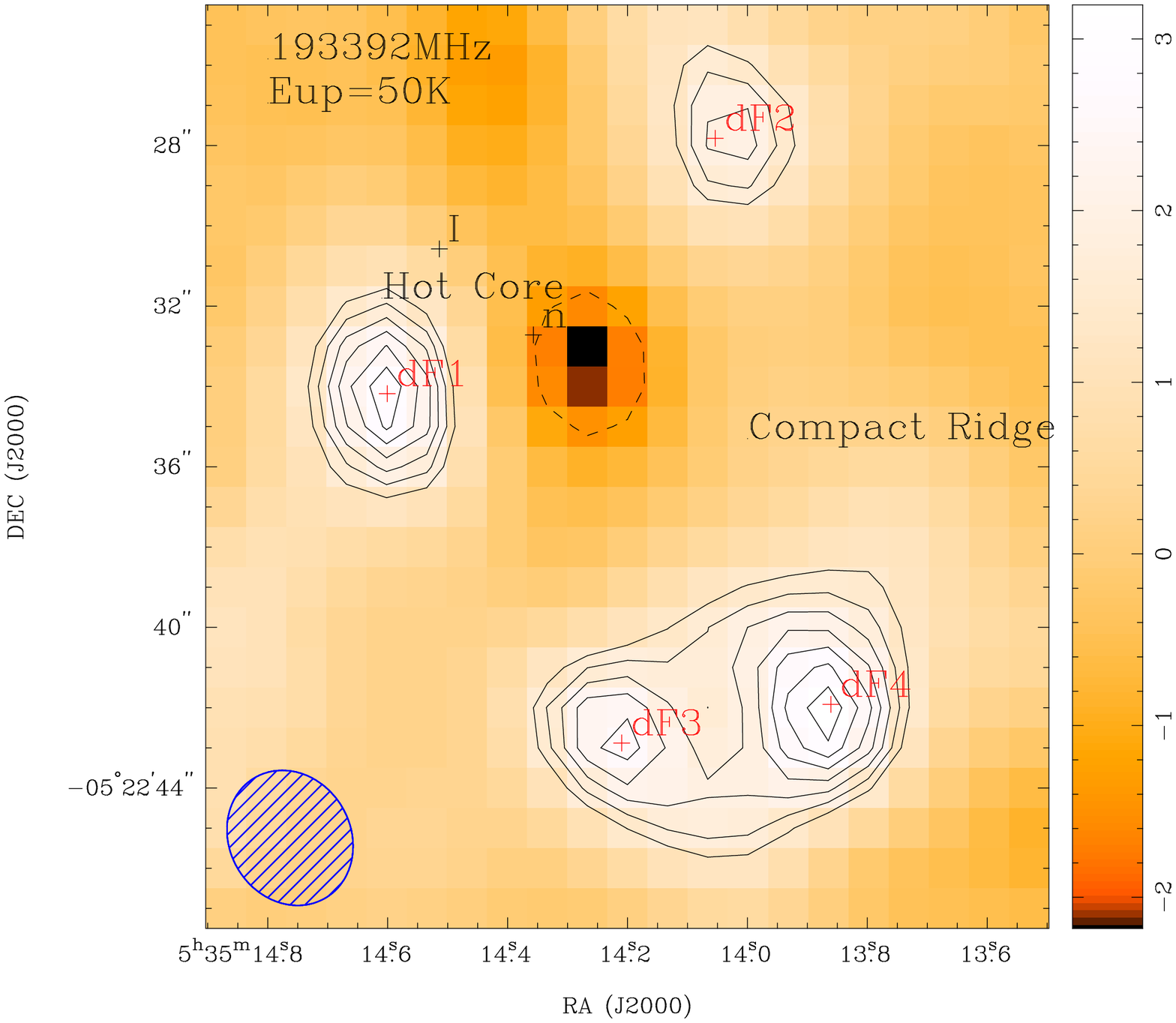}
\includegraphics[angle=270,width=8cm]{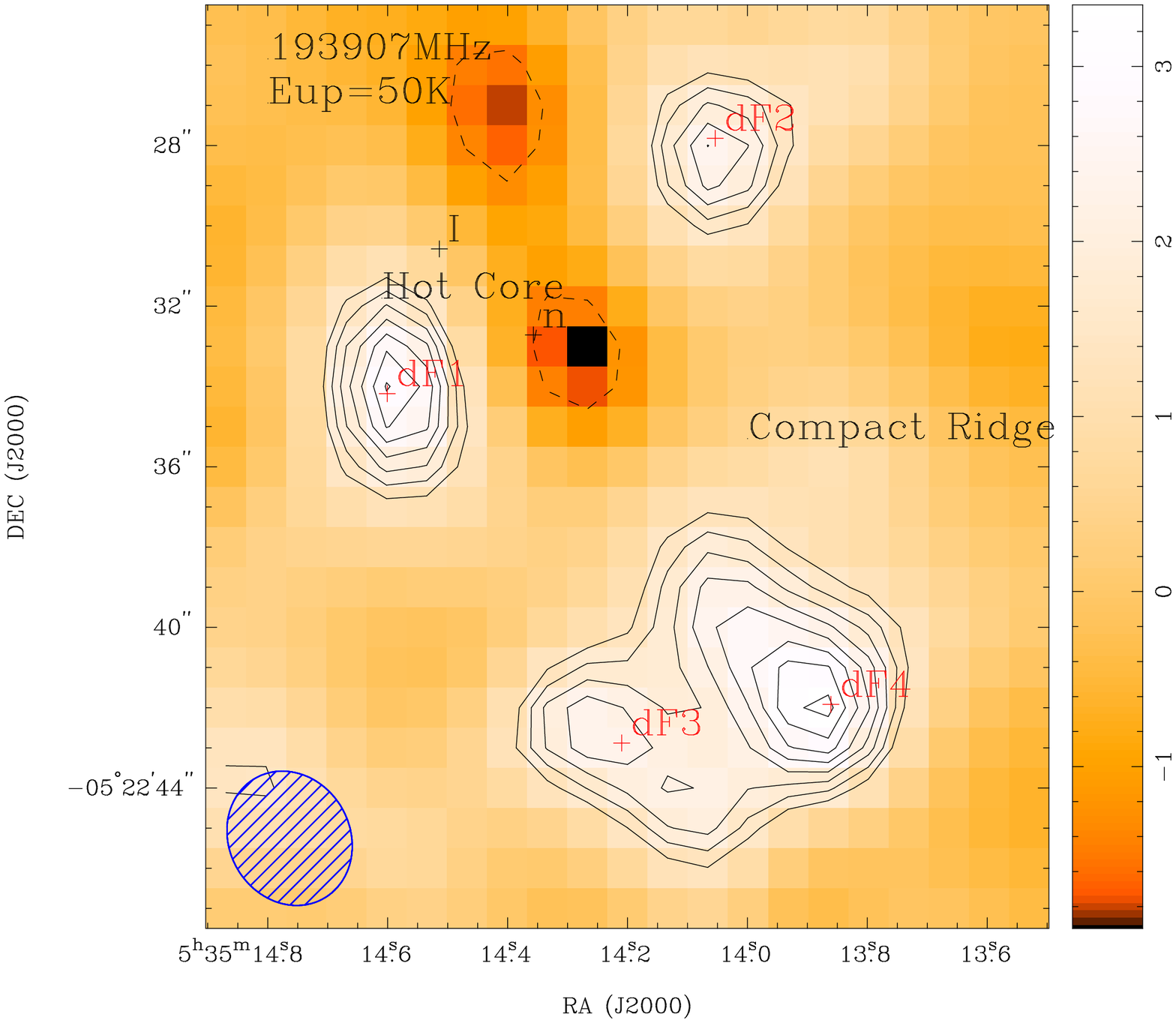}

\includegraphics[angle=270,width=8cm]{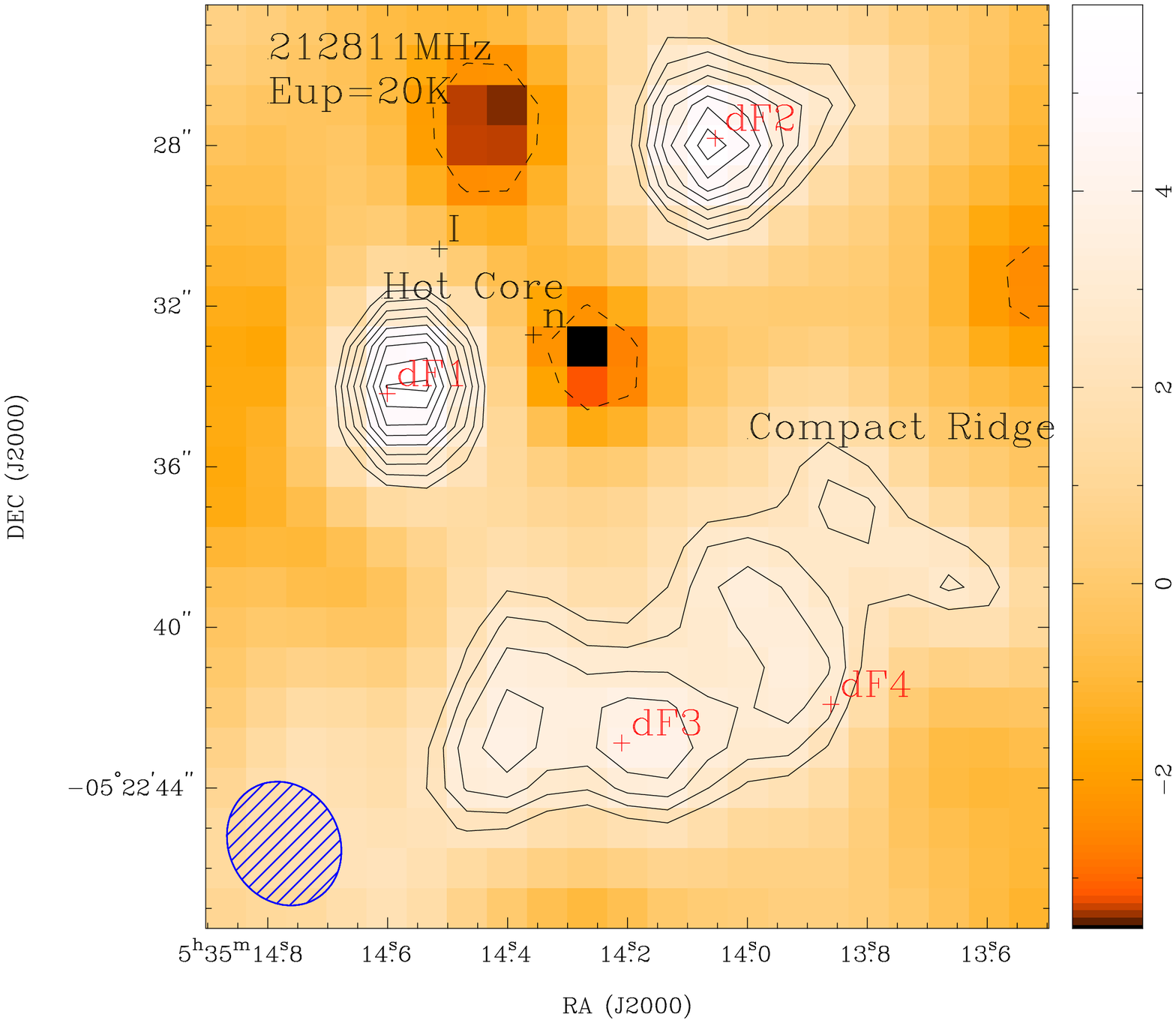}
\includegraphics[angle=270,width=8cm]{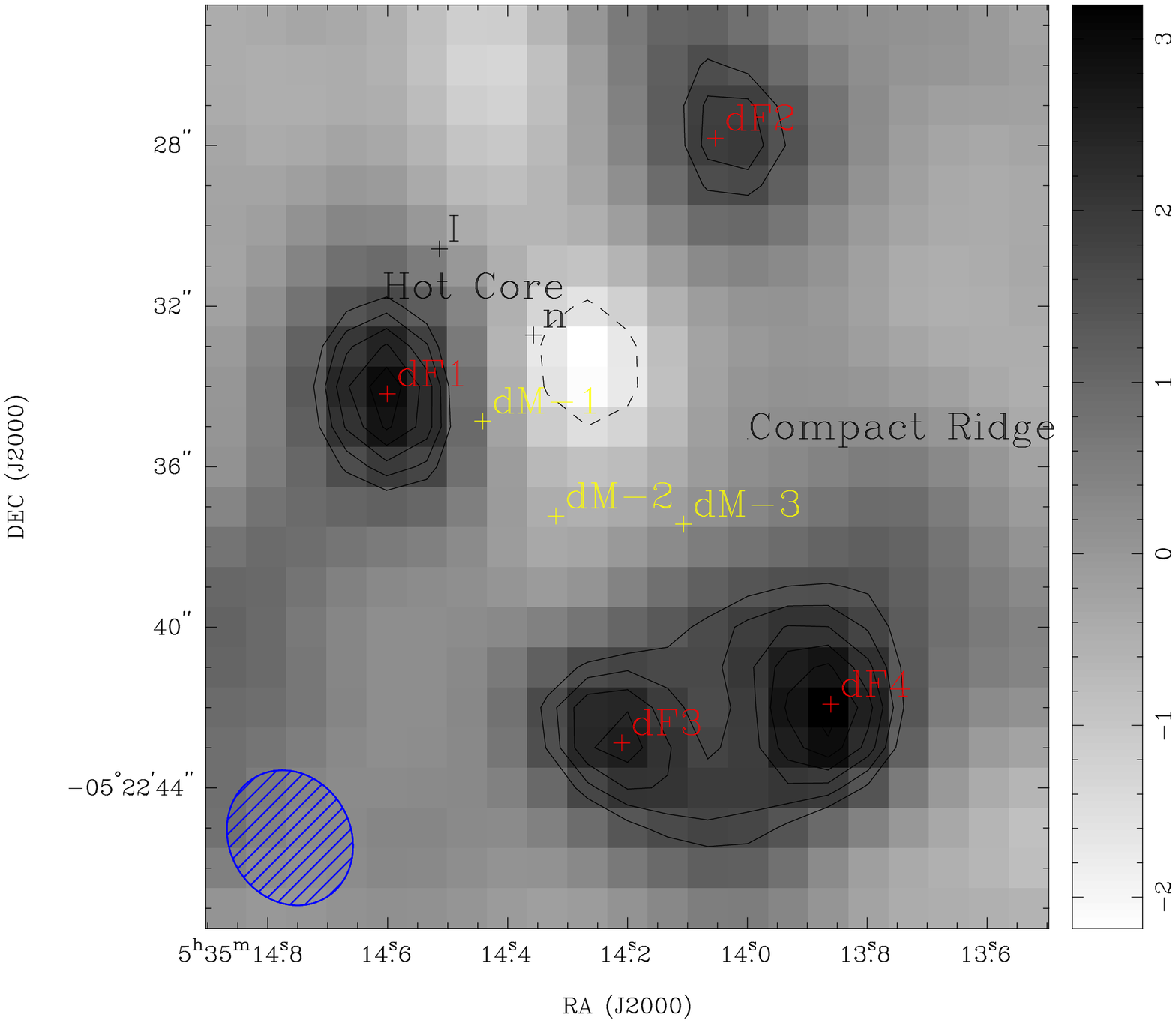}
\caption{Formaldehyde integrated emission maps from $v_{LSR}$=6.5~km~s$^{-1}$ to $v_{LSR}$=10.0~km~s$^{-1}$. Red crosses indicate the position of the regions dF1 to dF4. Black crosses indicate the positions of the radio source I ($\alpha_{J2000}$ = 05$^{h}$35$^{m}$14$\fs$5141, $\delta_{J2000}$ = -05$\degr$22$\arcmin$30$\farcs$575) and the IR source n ($\alpha_{J2000}$ = 05$^{h}$35$^{m}$14$\fs$3571, $\delta_{J2000}$ = -05$\degr$22$\arcmin$32$\farcs$719) \citep{Goddi:2011}. {\em Top:} HDCO emission at 193391.6~MHz (left panel) and 193907.5~MHz (right panel). {\em Bottom left:} H$_2^{13}$CO emission at  212811.2~MHz. The first contour is at 5$\sigma$ and the level step at 1$\sigma$ (where $\sigma$=0.27 and 0.44~Jy~beam$^{-1}$~km~s$^{-1}$ for HDCO and H$_2^{13}$CO, respectively). {\em Bottom right:} same as the top left panel, expect for deuterated methanol emission peaks (in yellow) identified by \citet{Peng:2012} are indicated.\label{fg2}}
\end{figure*}

\begin{figure*}
\centering
\includegraphics[angle=270,width=10cm]{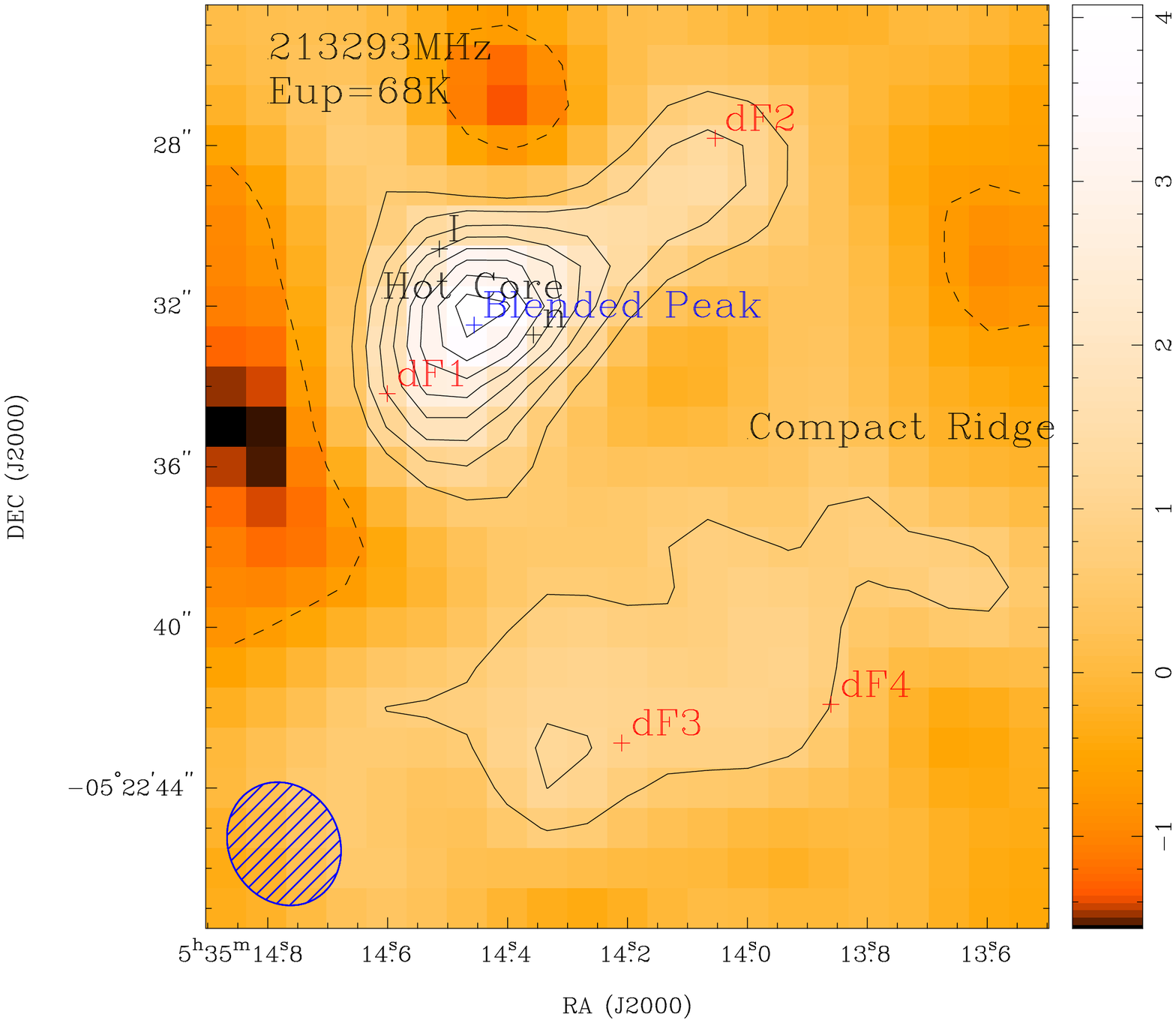}
\includegraphics[angle=0,width=7cm]{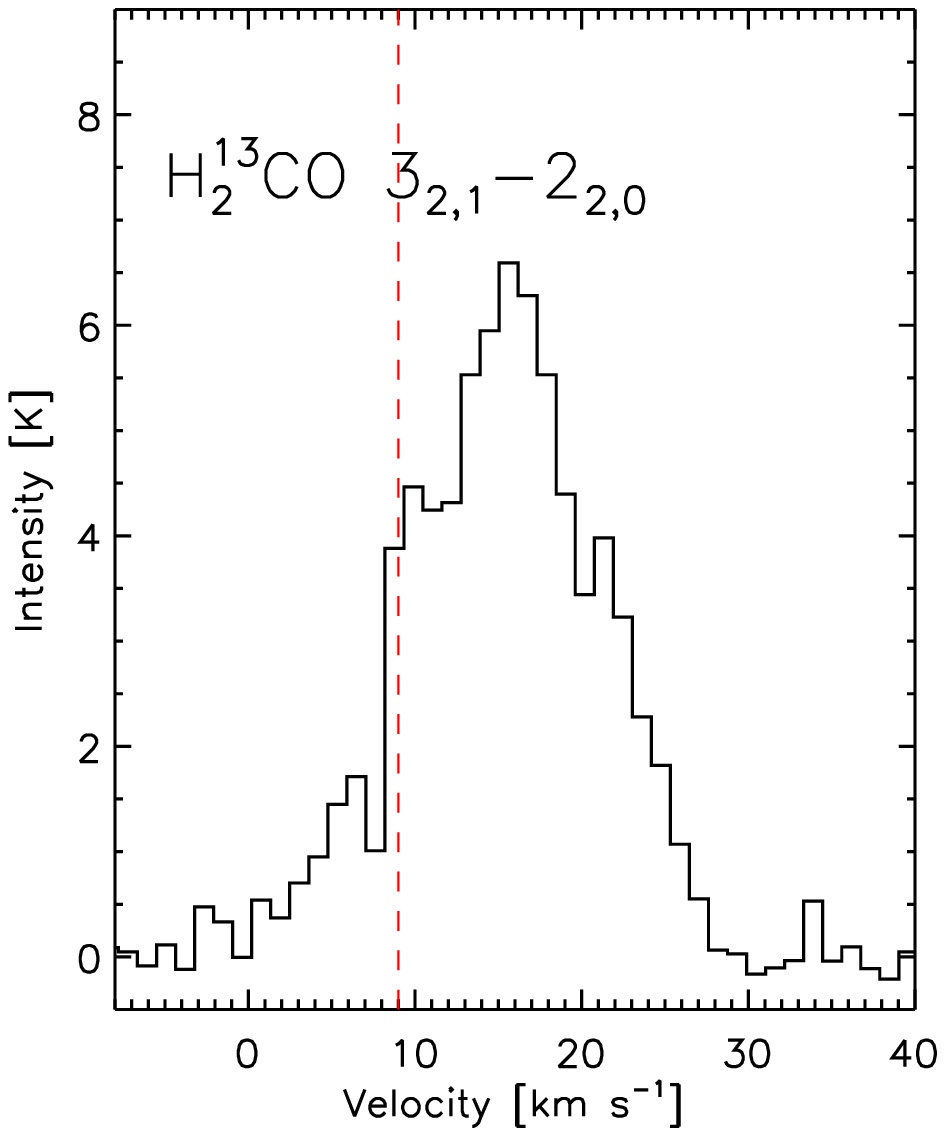}
\caption{ {\em Top:}  H$_2^{13}$CO integrated emission map at  213293.6~MHz from $v_{LSR}$=6.5~km~s$^{-1}$ to $v_{LSR}$=10.0~km~s$^{-1}$. 
Red crosses indicate the position of the regions dF1 to dF4 while black crosses indicate the positions of the radio source I  and the IR source n (see Fig.~\ref{fg2}). The first contour is at 3$\sigma$ and the level step at 2$\sigma$ (where $\sigma$=0.27~Jy~beam$^{-1}$~km~s$^{-1}$). The emission of this H$_2$$^{13}$CO transition is strongly contaminated by the emission from an unidentified species toward the Hot Core region marked by a blue cross and labelled "Blended Peak". {\em Bottom:} H$_2^{13}$CO spectrum observed towards the "Blended Peak" position (blue cross). The red dashed line indicates a LSR velocity of 9.0~km~s$^{-1}$.\label{fg3}}
\end{figure*}

\begin{figure*}
\includegraphics[angle=0,width=\textwidth]{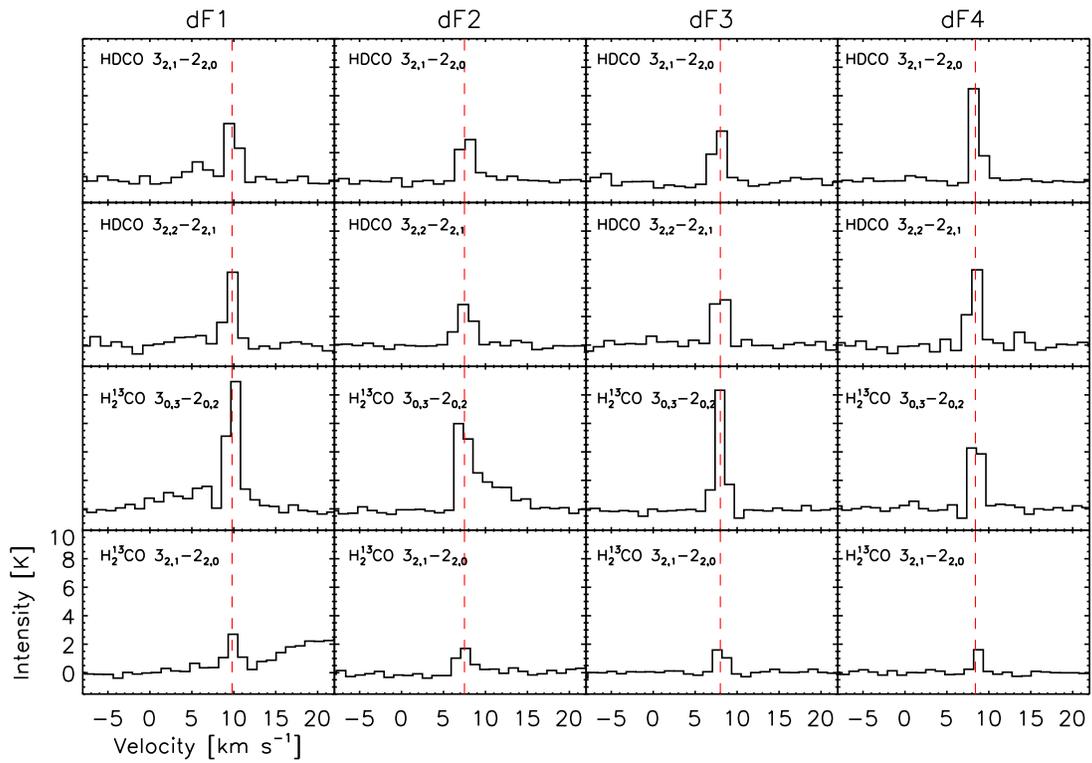}
\caption{HDCO and H$_2^{13}$CO spectra observed towards dF1, dF2, dF3 and dF4. The red dashed line indicates the $v_{LSR}$ towards each peak (i.e 9.8, 7.5, 8.0 and 8.4~km~s$^{-1}$, respectively).\label{fg4}}
\end{figure*}

\begin{figure*}
\centering
\includegraphics[angle=270,width=10.cm]{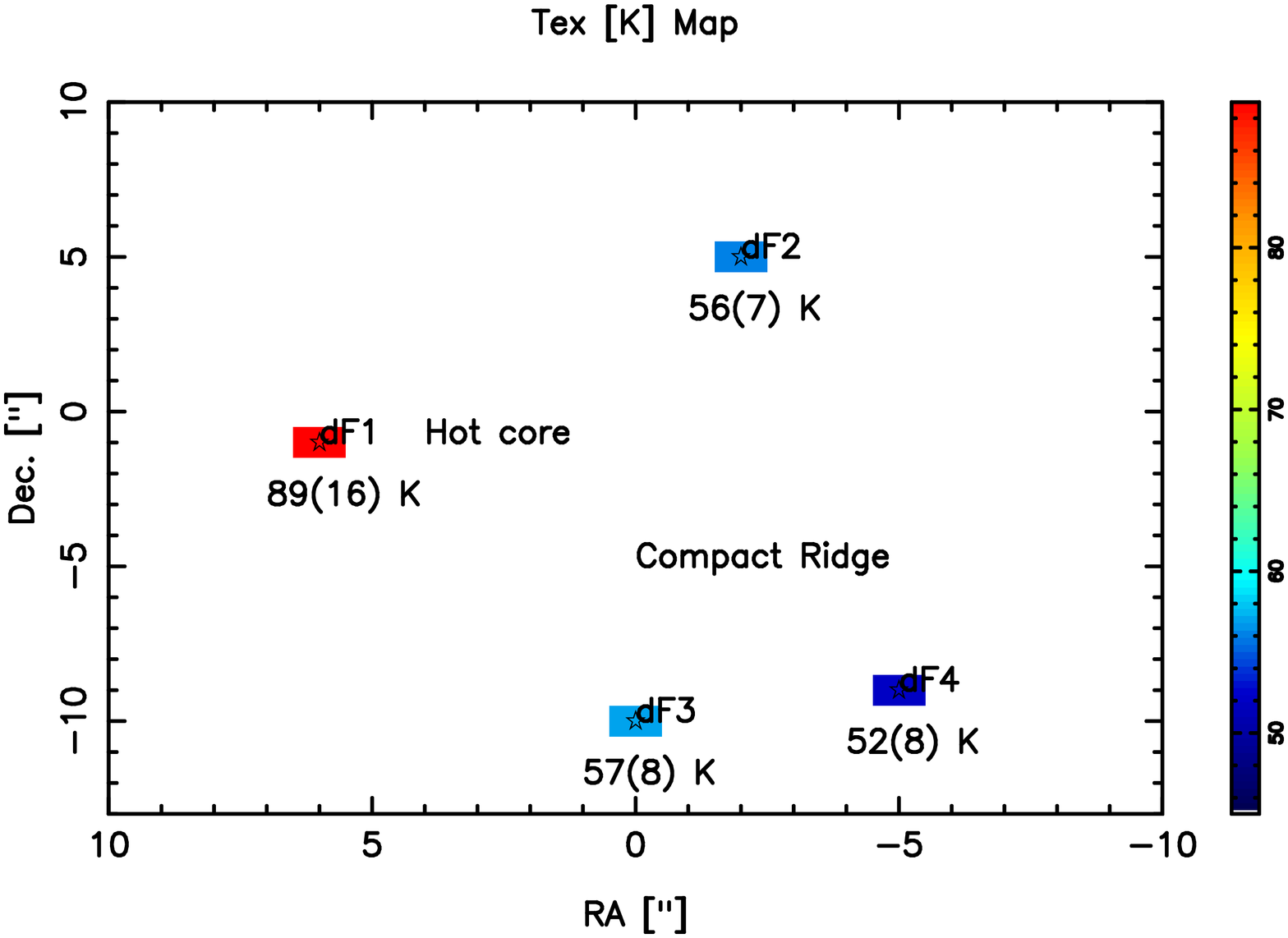}

\includegraphics[angle=270,width=10.cm]{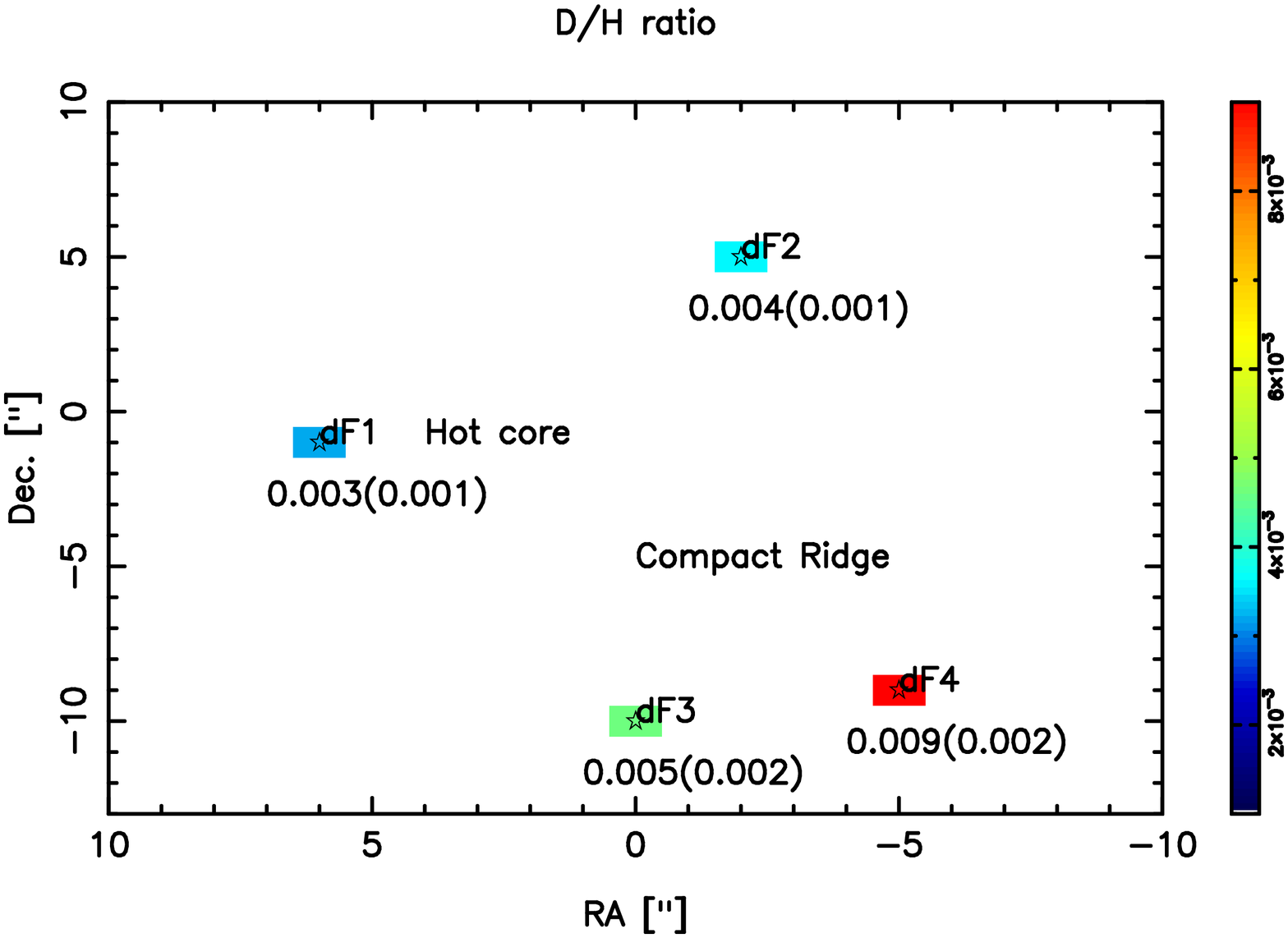}
\caption{{\em Top:} H$_2^{13}$CO excitation temperature map toward Orion-KL. {\em Bottom:} D/H ratio for formaldehyde within Orion--KL. \label{fg5}}
\end{figure*}

A recent study by \citet{Peng:2012} using the Plateau de Bure interferometer derived the D/H ratios of methanol (CH$_2$DOH and CH$_3$OD) at comparable resolution to our SMA observations. A notable result is that the spatial distribution of formaldehyde emission (HDCO and H$_2$$^{13}$CO, see Fig.~\ref{fg2}) clearly differs from that of (deuterated) methanol \citep[see Fig. 3 of][]{Peng:2012}. The methanol distribution follows a V-shaped structure linking the Hot Core component to the Compact Ridge \citep{Peng:2012} while formaldehyde exhibits a spatial structure which appears to surround this molecular V-shaped structure on four sides. This difference is illustrated in Figure~\ref{fg2} (bottom right panel) in which the deuterated methanol emission peaks identified by \citet{Peng:2012} are indicated on the HDCO integrated emission map at at 193391.6~MHz.  Furthermore, the D/H ratio for methanol derived by \citet{Peng:2012} lies in the range 0.8--1.3$\times$10$^{-3}$, which is up to an order of magnitude (within the uncertainties) lower than the D/H ratio for formaldehyde derived in this study (see Table~\ref{tab2}).

\subsection{Origin of formaldehyde emission}

All the nearby well--characterized sources (i.e. Hot Core, Compact Ridge) are not coincident with formaldehyde emission. In fact, it is possible and even likely that we are seeing quiescent material surrounding the hotter sources, such as the well--known quiescent ridge in Orion--KL. Incidentally, the spatial difference between the formaldehyde and methanol emission has already been observed toward others sources, such as the Horsehead Photon-Dominated Region (PDR) and its associated dense core \citep{Guzman:2013} and the Orion Bar PDR \citep{Leurini:2006,Leurini:2010}. 
In their respective analyses, the authors have shown that the UV radiation field affects the gas--phase abundance of formaldehyde and methanol. More specifically, the H$_2$CO/CH$_3$OH abundance ratio is expected to be higher in high FUV illuminated environments in comparison to denser and less illuminated regions. 
In this light, the entire Orion--KL quiescent ridge gas is a face--on PDR \citep[see][]{Lerate:2008,Persson:2007,Irvine:1987}, therefore photodissociation of gas-phase methanol into formaldehyde might occur \citep[e.g. see,][]{Le-Teuff:2000,Leurini:2006,Leurini:2010}.
In this context we determine the abundance ratio of H$_2$CO/CH$_3$OH in the formaldehyde clumps. Comparison to ratios measured towards other PDRs would then illuminate whether the Orion PDR is influencing the local chemistry of dense material surrounding the hot embedded cores. 

\begin{figure*}
\includegraphics[angle=0,width=\textwidth]{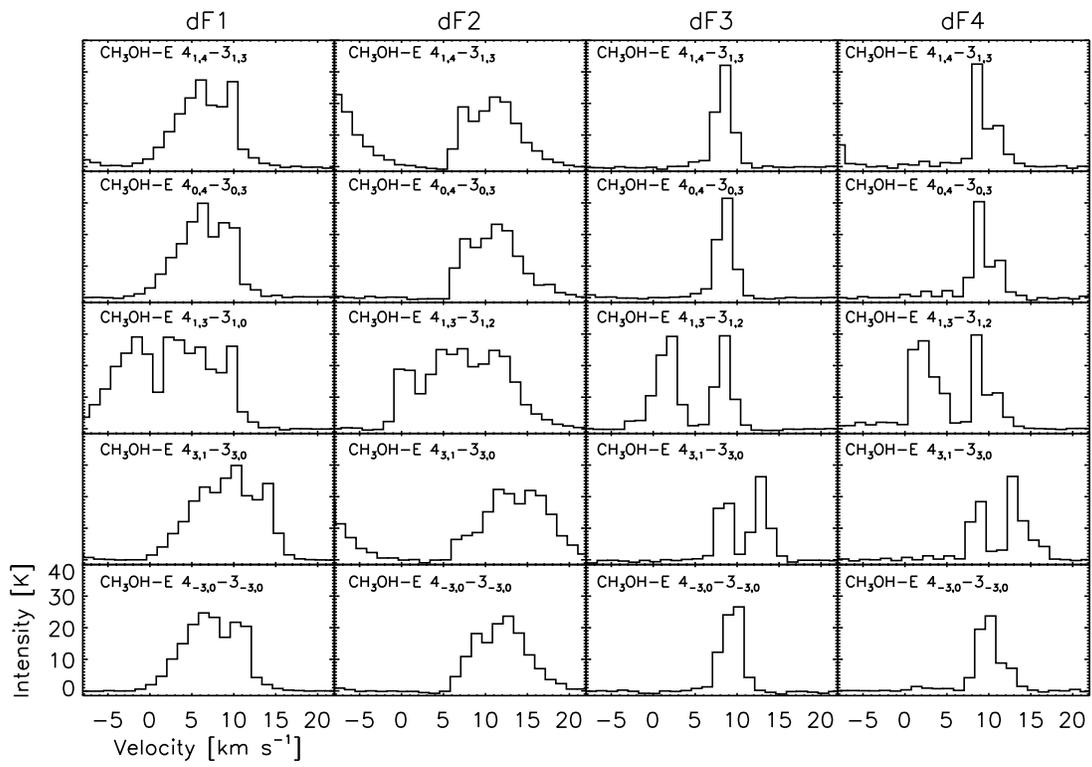}
\caption{ CH$_{3}$OH--E spectra observed towards dF1, dF2, dF3 and dF4.\label{fg6}}
\end{figure*}

The determination of the methanol column density (see Eq.~2) relies on the assumptions that the CH$_3$OH--E transitions listed in Table~\ref{tab3} are optically thin and, that the rotational temperature is the same as that of H$_2$$^{13}$CO. It is important to note that toward the dF1 and dF2 emission peaks the CH$_3$OH--E lines are strongly blended (see Figure~\ref{fg6}), preventing us from deriving the column density of methanol at those positions. Regarding the dF3 and dF4 regions, the resulting H$_2$CO/CH$_3$OH abundance ratio lies in the range 3.5--4.5, that might suggest a slight decrease in the methanol abundance in the quiescent ridge. However, our derived ratio is lower by at least a  factor of 3--400 in comparison to that estimated by \citet{Leurini:2010} toward the Orion Bar, suggesting that the production of formaldehyde in Orion--KL through photodissociation of gas-phase methanol is not the main formation route.
Finally, it is noticeable that our estimated H$_2$CO/CH$_3$OH abundance ratio is more consistent with that found in hot--corino regions (see \citealp{Maret:2004,Maret:2005} and \citealp{Guzman:2013} for a review).
This finding leads us to argue that formaldehyde emission in Orion--KL is more likely probing the edge of the hot gas.

\subsection{Implications: on the formaldehyde formation}

Formaldehyde and methanol are believed to be formed on ice grain mantles during the pre-stellar cold phase (T$<$50~K) from successive hydrogenation reactions of H-- or D-- atoms with CO \citep[see e.g.][]{Charnley:1997,Cazaux:2011,Taquet:2012a,Watanabe:2002}:
\begin{equation}
\rm
{CO} \rightarrow HCO \rightarrow {H}_2{CO} \rightarrow {CH}_3{O} \rightarrow {CH}_3{OH} 
\label{eq3}
\end{equation}
and
\begin{equation}
\rm
{CO} \rightarrow {DCO} \rightarrow {HDCO} \rightarrow {CH}_2{DO} \rightarrow {CH}_2{DOH}
\label{eq4}
\end{equation}
Laboratory experiments have shown that the sequential CO hydrogenation process (i.e. reaction~\ref{eq3}) efficiently form both formaldehyde and methanol at low temperatures \citep[$\le$20~K, see][]{Fuchs:2009,Watanabe:2004,Hidaka:2004}. Thus, if methanol and formaldehyde both purely originate from ice grain mantles, the measured D/H ratio should be similar (within a factor 2) for both molecules and, in particular, these molecules should trace similar environments while desorbed as they share a common origin. 
This is inconsistent with our observations (see Section~\ref{sec:anti}) and thus strongly argues in favor of differences in the chemical formation pathways of these deuterated molecules.
Thus CH$_3$OH and H$_2$CO emission is probing different spatial regions within the central region of Orion KL that may not be coeval. We infer that a likely scenario is a HDCO production through hot gas-phase deuterium chemistry in the colder gas associated with Orion--KL, as suggested by \citet{Loren:1985}. More specifically, at {\bf relatively} high temperature \citep[T$\sim$70K--100K,][]{Oberg:2012,Wootten:1987}, such as that derived in this study (see Figure~\ref{fg5}), reaction~\ref{eq5} is active due to its high exothermicity \citep[$\Delta$E of 654~K, see][]{Roueff:2013} and favors deuterium fractionation through reactions involving CH$_2$D$^{+}$ as a parent molecule \citep{Favre:2015}. 
\begin{equation}
\label{eq5}
\textnormal{CH}_3^{+} + \textnormal{HD} \leftrightarrows \textnormal{CH}_2\textnormal{D}^{+} + \textnormal{H}_2 + \textnormal{$\Delta$E}
\end{equation}
In that context, reaction~\ref{eq5} can then lead to the formation of HDCO and H$_2$CO \citep[see further details in][]{Loren:1985}. Incidentally, the temperature of this colder region associated with Orion-KL is not sufficient enough to desorb CH$_3$OH from the surface of icy grain mantles \citep[T$\ge$120--160~K see,][]{Collings:2004a,Green:2009}.
However, in the inner hottest part of Orion--KL where methanol emits, most of the formaldehyde might have been converted into methanol by grain surface reactions \citep{Watanabe:2002} during a colder phase prior to the explosive event. 
Then, upon heating methanol is released from grain surfaces into  gas phase along with little  formaldehyde. We stress that the release of methanol into gas--phase in this region is mainly due to internal radiation heating (i.e outflow, shock, IR sources) but not heat from the PDR (see Section 4.2).
In this instance, the D/H ratio and the spatial distribution of methanol should differ from the one of formaldehyde.

%
\section{Conclusions}
\label{sec:conc}

We have characterized, for the first time, the spatial distribution of deuterated formaldehyde (HDCO) toward Orion--KL with an angular resolution of 3.4\arcsec $\times$ 3.0\arcsec using the SMA. The new SMA broadband capability of the 230~GHz band allowed simultaneous observation of both  HDCO and H$_2$$^{13}$CO within relatively sparse spectra. Four main emission peaks, labelled in this study dF1 to dF4, are identified in the formaldehyde emission image. A salient result is that the spatial distribution the emission of the HDCO and H$_2$$^{13}$CO emission differ from that of CH$_3$OH and its deuterated isotopologues.
The emission maps show that methanol emits in the inner part of Orion--KL, while formaldehyde is found in the colder gas enveloping the CH$_3$OH emission on four sides: one located left of the Hot Core (dF1), one toward the north of the Compact Ridge (dF2) and two in the quiescent ridge (dF3 and dF4).  In addition, we derive a HDCO/H$_2$CO ratio which lies in the range of  0.003 to 0.009 within the region; this value is an order of magnitude higher than the D/H measured for methanol. We suggest that the high measured D/H ratio for formaldehyde and its spatial distribution in Orion--KL are a combination of gas phase chemistry in the cooler gas  associated with Orion--KL and, a conversion of formaldehyde into methanol on ice grain mantles in the earlier history of the inner part.

%
\acknowledgments
Support for this work was provided by NASA (Herschel OT funding) through an award issued by JPL/Caltech. This paper makes use of SMA data. C.F. thanks Tzu--Cheng Peng for helpful discussions.

{\it Facilities:} \facility{SMA}

%
\bibliographystyle{apj}

\begin{thebibliography}{76}
\expandafter\ifx\csname natexlab\endcsname\relax\def\natexlab#1{#1}\fi

\bibitem[{{Bally} {et~al.}(2011){Bally}, {Cunningham}, {Moeckel}, {Burton},
  {Smith}, {Frank}, \& {Nordlund}}]{Bally:2011}
{Bally}, J., {Cunningham}, N.~J., {Moeckel}, N., {et~al.} 2011, \apj, 727, 113

\bibitem[{{Bergin} {et~al.}(1994){Bergin}, {Goldsmith}, {Snell}, \&
  {Ungerechts}}]{Bergin:1994}
{Bergin}, E.~A., {Goldsmith}, P.~F., {Snell}, R.~L., \& {Ungerechts}, H. 1994,
  \apj, 431, 674

\bibitem[{{Bergin} {et~al.}(2010){Bergin}, {Phillips}, {Comito}, {Crockett},
  {Lis}, {Schilke}, {Wang}, {Bell}, {Blake}, {Bumble}, {Caux}, {Cabrit},
  {Ceccarelli}, {Cernicharo}, {Daniel}, {de Graauw}, {Dubernet},
  {Emprechtinger}, {Encrenaz}, {Falgarone}, {Gerin}, {Giesen}, {Goicoechea},
  {Goldsmith}, {Gupta}, {Hartogh}, {Helmich}, {Herbst}, {Joblin}, {Johnstone},
  {Kawamura}, {Langer}, {Latter}, {Lord}, {Maret}, {Martin}, {Melnick},
  {Menten}, {Morris}, {M{\"u}ller}, {Murphy}, {Neufeld}, {Ossenkopf}, {Pagani},
  {Pearson}, {P{\'e}rault}, {Plume}, {Roelfsema}, {Qin}, {Salez}, {Schlemmer},
  {Stutzki}, {Tielens}, {Trappe}, {van der Tak}, {Vastel}, {Yorke}, {Yu}, \&
  {Zmuidzinas}}]{Bergin:2010}
{Bergin}, E.~A., {Phillips}, T.~G., {Comito}, C., {et~al.} 2010, \aap, 521, L20

\bibitem[{{Beuther} {et~al.}(2005){Beuther}, {Zhang}, {Greenhill}, {Reid},
  {Wilner}, {Keto}, {Shinnaga}, {Ho}, {Moran}, {Liu}, \&
  {Chang}}]{Beuther:2005}
{Beuther}, H., {Zhang}, Q., {Greenhill}, L.~J., {et~al.} 2005, \apj, 632, 355

\bibitem[{{Beuther} {et~al.}(2006){Beuther}, {Zhang}, {Reid}, {Hunter},
  {Gurwell}, {Wilner}, {Zhao}, {Shinnaga}, {Keto}, {Ho}, {Moran}, \&
  {Liu}}]{Beuther:2006}
{Beuther}, H., {Zhang}, Q., {Reid}, M.~J., {et~al.} 2006, \apj, 636, 323

\bibitem[{{Blake} {et~al.}(1987){Blake}, {Sutton}, {Masson}, \&
  {Phillips}}]{Blake:1987}
{Blake}, G.~A., {Sutton}, E.~C., {Masson}, C.~R., \& {Phillips}, T.~G. 1987,
  \apj, 315, 621

\bibitem[{{Bocquet} {et~al.}(1999){Bocquet}, {Demaison}, {Cosl{\'e}ou},
  {Friedrich}, {Margul{\`e}s}, {Macholl}, {M{\"a}der}, {Beaky}, \&
  {Winnewisser}}]{Bocquet:1999}
{Bocquet}, R., {Demaison}, J., {Cosl{\'e}ou}, J., {et~al.} 1999, Journal of
  Molecular Spectroscopy, 195, 345

\bibitem[{{Brouillet} {et~al.}(2013){Brouillet}, {Despois}, {Baudry}, {Peng},
  {Favre}, {Wootten}, {Remijan}, {Wilson}, {Combes}, \&
  {Wlodarczak}}]{Brouillet:2013}
{Brouillet}, N., {Despois}, D., {Baudry}, A., {et~al.} 2013, \aap, 550, A46

\bibitem[{{Cazaux} {et~al.}(2011){Cazaux}, {Caselli}, \&
  {Spaans}}]{Cazaux:2011}
{Cazaux}, S., {Caselli}, P., \& {Spaans}, M. 2011, \apjl, 741, L34

\bibitem[{{Charnley} {et~al.}(1997){Charnley}, {Tielens}, \&
  {Rodgers}}]{Charnley:1997}
{Charnley}, S.~B., {Tielens}, A.~G.~G.~M., \& {Rodgers}, S.~D. 1997, \apjl,
  482, L203

\bibitem[{{Collings} {et~al.}(2004){Collings}, {Anderson}, {Chen}, {Dever},
  {Viti}, {Williams}, \& {McCoustra}}]{Collings:2004a}
{Collings}, M.~P., {Anderson}, M.~A., {Chen}, R., {et~al.} 2004, \mnras, 354,
  1133

\bibitem[{{Comito} {et~al.}(2005){Comito}, {Schilke}, {Phillips}, {Lis},
  {Motte}, \& {Mehringer}}]{Comito:2005}
{Comito}, C., {Schilke}, P., {Phillips}, T.~G., {et~al.} 2005, \apjs, 156, 127

\bibitem[{{Crockett} {et~al.}(2010){Crockett}, {Bergin}, {Wang}, {Lis}, {Bell},
  {Blake}, {Boogert}, {Bumble}, {Cabrit}, {Caux}, {Ceccarelli}, {Cernicharo},
  {Comito}, {Daniel}, {Dubernet}, {Emprechtinger}, {Encrenaz}, {Falgarone},
  {Gerin}, {Giesen}, {Goicoechea}, {Goldsmith}, {Gupta}, {G{\"u}sten},
  {Hartogh}, {Helmich}, {Herbst}, {Honingh}, {Joblin}, {Johnstone}, {Karpov},
  {Kawamura}, {Kooi}, {Krieg}, {Langer}, {Latter}, {Lord}, {Maret}, {Martin},
  {Melnick}, {Menten}, {Morris}, {M{\"u}ller}, {Murphy}, {Neufeld},
  {Ossenkopf}, {Pearson}, {P{\'e}rault}, {Phillips}, {Plume}, {Qin},
  {Roelfsema}, {Schieder}, {Schilke}, {Schlemmer}, {Stutzki}, {van der Tak},
  {Tielens}, {Trappe}, {Vastel}, {Yorke}, {Yu}, \&
  {Zmuidzinas}}]{Crockett:2010}
{Crockett}, N.~R., {Bergin}, E.~A., {Wang}, S., {et~al.} 2010, \aap, 521, L21+

\bibitem[{{Crockett} {et~al.}(2014){Crockett}, {Bergin}, {Neill}, {Favre},
  {Schilke}, {Lis}, {Bell}, {Blake}, {Cernicharo}, {Emprechtinger},
  {Esplugues}, {Gupta}, {Kleshcheva}, {Lord}, {Marcelino}, {McGuire},
  {Pearson}, {Phillips}, {Plume}, {van der Tak}, {Tercero}, \&
  {Yu}}]{Crockett:2014}
{Crockett}, N.~R., {Bergin}, E.~A., {Neill}, J.~L., {et~al.} 2014, \apj, 787,
  112

\bibitem[{{Esplugues} {et~al.}(2013){Esplugues}, {Tercero}, {Cernicharo},
  {Goicoechea}, {Palau}, {Marcelino}, \& {Bell}}]{Esplugues:2013}
{Esplugues}, G.~B., {Tercero}, B., {Cernicharo}, J., {et~al.} 2013, \aap, 556,
  A143

\bibitem[{{Favre} {et~al.}(2015){Favre}, {Bergin}, {Cleeves}, {Hersant}, {Qi},
  \& {Aikawa}}]{Favre:2015}
{Favre}, C., {Bergin}, E.~A., {Cleeves}, L.~I., {et~al.} 2015, \apjl, 802, L23

\bibitem[{{Favre} {et~al.}(2011{\natexlab{a}}){Favre}, {Despois}, {Brouillet},
  {Baudry}, {Combes}, {Gu{\'e}lin}, {Wootten}, \& {Wlodarczak}}]{Favre:2011}
{Favre}, C., {Despois}, D., {Brouillet}, N., {et~al.} 2011{\natexlab{a}}, \aap,
  532, A32

\bibitem[{{Favre} {et~al.}(2011{\natexlab{b}}){Favre}, {Wootten}, {Remijan},
  {Brouillet}, {Wilson}, {Despois}, \& {Baudry}}]{Favre:2011a}
{Favre}, C., {Wootten}, H.~A., {Remijan}, A.~J., {et~al.} 2011{\natexlab{b}},
  \apjl, 739, L12

\bibitem[{{Favre} {et~al.}(2014){Favre}, {Carvajal}, {Field}, {Jorgensen},
  {Bisschop}, {Brouillet}, {Despois}, {Baudry}, {Kleiner}, {Bergin},
  {Crockett}, {Neill}, {Margules}, {Huet}, \& {Demaison}}]{Favre:2014a}
{Favre}, C., {Carvajal}, M., {Field}, D., {et~al.} 2014, \apjs, 215, 25

\bibitem[{{Friedel} \& {Snyder}(2008)}]{Friedel:2008}
{Friedel}, D.~N., \& {Snyder}, L.~E. 2008, \apj, 672, 962

\bibitem[{{Friedel} \& {Widicus Weaver}(2012)}]{Friedel:2012}
{Friedel}, D.~N., \& {Widicus Weaver}, S.~L. 2012, \apjs, 201, 17

\bibitem[{{Fuchs} {et~al.}(2009){Fuchs}, {Cuppen}, {Ioppolo}, {Romanzin},
  {Bisschop}, {Andersson}, {van Dishoeck}, \& {Linnartz}}]{Fuchs:2009}
{Fuchs}, G.~W., {Cuppen}, H.~M., {Ioppolo}, S., {et~al.} 2009, \aap, 505, 629

\bibitem[{{Goddi} {et~al.}(2011{\natexlab{a}}){Goddi}, {Greenhill},
  {Humphreys}, {Chandler}, \& {Matthews}}]{Goddi:2011a}
{Goddi}, C., {Greenhill}, L.~J., {Humphreys}, E.~M.~L., {Chandler}, C.~J., \&
  {Matthews}, L.~D. 2011{\natexlab{a}}, \apjl, 739, L13

\bibitem[{{Goddi} {et~al.}(2011{\natexlab{b}}){Goddi}, {Humphreys},
  {Greenhill}, {Chandler}, \& {Matthews}}]{Goddi:2011}
{Goddi}, C., {Humphreys}, E.~M.~L., {Greenhill}, L.~J., {Chandler}, C.~J., \&
  {Matthews}, L.~D. 2011{\natexlab{b}}, \apj, 728, 15

\bibitem[{{Goldsmith} \& {Langer}(1999)}]{Goldsmith:1999}
{Goldsmith}, P.~F., \& {Langer}, W.~D. 1999, \apj, 517, 209

\bibitem[{{G{\'o}mez} {et~al.}(2008){G{\'o}mez}, {Rodr{\'{\i}}guez}, {Loinard},
  {Lizano}, {Allen}, {Poveda}, \& {Menten}}]{Gomez:2008}
{G{\'o}mez}, L., {Rodr{\'{\i}}guez}, L.~F., {Loinard}, L., {et~al.} 2008, \apj,
  685, 333

\bibitem[{{G{\'o}mez} {et~al.}(2005){G{\'o}mez}, {Rodr{\'{\i}}guez}, {Loinard},
  {Lizano}, {Poveda}, \& {Allen}}]{Gomez:2005}
---. 2005, \apj, 635, 1166

\bibitem[{{Green} {et~al.}(2009){Green}, {Bolina}, {Chen}, {Collings}, {Brown},
  \& {McCoustra}}]{Green:2009}
{Green}, S.~D., {Bolina}, A.~S., {Chen}, R., {et~al.} 2009, \mnras, 398, 357

\bibitem[{{Gu{\'e}lin} {et~al.}(2008){Gu{\'e}lin}, {Brouillet}, {Cernicharo},
  {Combes}, \& {Wooten}}]{Guelin:2008}
{Gu{\'e}lin}, M., {Brouillet}, N., {Cernicharo}, J., {Combes}, F., \& {Wooten},
  A. 2008, \apss, 313, 45

\bibitem[{{Guzm{\'a}n} {et~al.}(2013){Guzm{\'a}n}, {Goicoechea}, {Pety},
  {Gratier}, {Gerin}, {Roueff}, {Le Petit}, {Le Bourlot}, \&
  {Faure}}]{Guzman:2013}
{Guzm{\'a}n}, V.~V., {Goicoechea}, J.~R., {Pety}, J., {et~al.} 2013, \aap, 560,
  A73

\bibitem[{{Hidaka} {et~al.}(2004){Hidaka}, {Watanabe}, {Shiraki}, {Nagaoka}, \&
  {Kouchi}}]{Hidaka:2004}
{Hidaka}, H., {Watanabe}, N., {Shiraki}, T., {Nagaoka}, A., \& {Kouchi}, A.
  2004, \apj, 614, 1124

\bibitem[{{Ho} {et~al.}(2004){Ho}, {Moran}, \& {Lo}}]{Ho:2004}
{Ho}, P.~T.~P., {Moran}, J.~M., \& {Lo}, K.~Y. 2004, \apjl, 616, L1

\bibitem[{{Irvine} {et~al.}(1987){Irvine}, {Goldsmith}, \&
  {Hjalmarson}}]{Irvine:1987}
{Irvine}, W.~M., {Goldsmith}, P.~F., \& {Hjalmarson}, A. 1987, Astrophysics and
  Space Science Library, Vol. 134, {Chemical abundances in molecular clouds},
  ed. {D.~J.~Hollenbach \& H.~A.~Thronson Jr.}, 561--609

\bibitem[{{Jacq} {et~al.}(1990){Jacq}, {Walmsley}, {Henkel}, {Baudry},
  {Mauersberger}, \& {Jewell}}]{Jacq:1990}
{Jacq}, T., {Walmsley}, C.~M., {Henkel}, C., {et~al.} 1990, \aap, 228, 447

\bibitem[{{Jacq} {et~al.}(1993){Jacq}, {Walmsley}, {Mauersberger}, {Anderson},
  {Herbst}, \& {De Lucia}}]{Jacq:1993}
{Jacq}, T., {Walmsley}, C.~M., {Mauersberger}, R., {et~al.} 1993, \aap, 271,
  276

\bibitem[{{Johns} \& {McKellar}(1977)}]{Johns:1977}
{Johns}, J.~W.~C., \& {McKellar}, A.~R.~W. 1977, Journal of Molecular
  Spectroscopy, 64, 327

\bibitem[{{Kahane} {et~al.}(1984){Kahane}, {Lucas}, {Frerking}, {Langer}, \&
  {Encrenaz}}]{Kahane:1984}
{Kahane}, C., {Lucas}, R., {Frerking}, M.~A., {Langer}, W.~D., \& {Encrenaz},
  P. 1984, \aap, 137, 211

\bibitem[{{Le Teuff} {et~al.}(2000){Le Teuff}, {Millar}, \&
  {Markwick}}]{Le-Teuff:2000}
{Le Teuff}, Y.~H., {Millar}, T.~J., \& {Markwick}, A.~J. 2000, \aaps, 146, 157

\bibitem[{{Lerate} {et~al.}(2008){Lerate}, {Yates}, {Viti}, {Barlow},
  {Swinyard}, {White}, {Cernicharo}, \& {Goicoechea}}]{Lerate:2008}
{Lerate}, M.~R., {Yates}, J., {Viti}, S., {et~al.} 2008, \mnras, 387, 1660

\bibitem[{{Leurini} {et~al.}(2010){Leurini}, {Parise}, {Schilke}, {Pety}, \&
  {Rolffs}}]{Leurini:2010}
{Leurini}, S., {Parise}, B., {Schilke}, P., {Pety}, J., \& {Rolffs}, R. 2010,
  \aap, 511, A82

\bibitem[{{Leurini} {et~al.}(2006){Leurini}, {Rolffs}, {Thorwirth}, {Parise},
  {Schilke}, {Comito}, {Wyrowski}, {G{\"u}sten}, {Bergman}, {Menten}, \&
  {Nyman}}]{Leurini:2006}
{Leurini}, S., {Rolffs}, R., {Thorwirth}, S., {et~al.} 2006, \aap, 454, L47

\bibitem[{{Loren} \& {Wootten}(1985)}]{Loren:1985}
{Loren}, R.~B., \& {Wootten}, A. 1985, \apj, 299, 947

\bibitem[{{Maret} {et~al.}(2005){Maret}, {Ceccarelli}, {Tielens}, {Caux},
  {Lefloch}, {Faure}, {Castets}, \& {Flower}}]{Maret:2005}
{Maret}, S., {Ceccarelli}, C., {Tielens}, A.~G.~G.~M., {et~al.} 2005, \aap,
  442, 527

\bibitem[{{Maret} {et~al.}(2004){Maret}, {Ceccarelli}, {Caux}, {Tielens},
  {J{\o}rgensen}, {van Dishoeck}, {Bacmann}, {Castets}, {Lefloch}, {Loinard},
  {Parise}, \& {Sch{\"o}ier}}]{Maret:2004}
{Maret}, S., {Ceccarelli}, C., {Caux}, E., {et~al.} 2004, \aap, 416, 577

\bibitem[{{Mauersberger} {et~al.}(1988){Mauersberger}, {Henkel}, {Jacq}, \&
  {Walmsley}}]{Mauersberger:1988}
{Mauersberger}, R., {Henkel}, C., {Jacq}, T., \& {Walmsley}, C.~M. 1988, \aap,
  194, L1

\bibitem[{{Menten} {et~al.}(2007){Menten}, {Reid}, {Forbrich}, \&
  {Brunthaler}}]{Menten:2007}
{Menten}, K.~M., {Reid}, M.~J., {Forbrich}, J., \& {Brunthaler}, A. 2007, \aap,
  474, 515

\bibitem[{{M{\"u}ller} {et~al.}(2005){M{\"u}ller}, {Schl{\"o}der}, {Stutzki},
  \& {Winnewisser}}]{Muller:2005}
{M{\"u}ller}, H.~S.~P., {Schl{\"o}der}, F., {Stutzki}, J., \& {Winnewisser}, G.
  2005, Journal of Molecular Structure, 742, 215

\bibitem[{{M{\"u}ller} {et~al.}(2000){M{\"u}ller}, {Gendriesch},
  {Margul{\`e}s}, {Lewen}, {Winnewisser}, {Bocquet}, {Demaison}, {W{\"o}tzel},
  \& {M{\"a}der}}]{Muller:2000}
{M{\"u}ller}, H.~S.~P., {Gendriesch}, R., {Margul{\`e}s}, L., {et~al.} 2000,
  Physical Chemistry Chemical Physics (Incorporating Faraday Transactions), 2,
  3401

\bibitem[{{Neill} {et~al.}(2013{\natexlab{a}}){Neill}, {Crockett}, {Bergin},
  {Pearson}, \& {Xu}}]{Neill:2013a}
{Neill}, J.~L., {Crockett}, N.~R., {Bergin}, E.~A., {Pearson}, J.~C., \& {Xu},
  L.-H. 2013{\natexlab{a}}, \apj, 777, 85

\bibitem[{{Neill} {et~al.}(2013{\natexlab{b}}){Neill}, {Wang}, {Bergin},
  {Crockett}, {Favre}, {Plume}, \& {Melnick}}]{Neill:2013}
{Neill}, J.~L., {Wang}, S., {Bergin}, E.~A., {et~al.} 2013{\natexlab{b}}, \apj,
  770, 142

\bibitem[{{Nissen} {et~al.}(2012){Nissen}, {Cunningham}, {Gustafsson}, {Bally},
  {Lemaire}, {Favre}, \& {Field}}]{Nissen:2012}
{Nissen}, H.~D., {Cunningham}, N.~J., {Gustafsson}, M., {et~al.} 2012, \aap,
  540, A119

\bibitem[{{{\"O}berg} {et~al.}(2012){{\"O}berg}, {Qi}, {Wilner}, \&
  {Hogerheijde}}]{Oberg:2012}
{{\"O}berg}, K.~I., {Qi}, C., {Wilner}, D.~J., \& {Hogerheijde}, M.~R. 2012,
  \apj, 749, 162

\bibitem[{{Pagani} {et~al.}(1992){Pagani}, {Salez}, \& {Wannier}}]{Pagani:1992}
{Pagani}, L., {Salez}, M., \& {Wannier}, P.~G. 1992, \aap, 258, 479

\bibitem[{{Peng} {et~al.}(2012){Peng}, {Despois}, {Brouillet}, {Parise}, \&
  {Baudry}}]{Peng:2012}
{Peng}, T.-C., {Despois}, D., {Brouillet}, N., {Parise}, B., \& {Baudry}, A.
  2012, \aap, 543, A152

\bibitem[{{Peng} {et~al.}(2013){Peng}, {Despois}, {Brouillet}, {Baudry},
  {Favre}, {Remijan}, {Wootten}, {Wilson}, {Combes}, \&
  {Wlodarczak}}]{Peng:2013}
{Peng}, T.-C., {Despois}, D., {Brouillet}, N., {et~al.} 2013, \aap, 554, A78

\bibitem[{{Persson} {et~al.}(2007){Persson}, {Olofsson}, {Koning}, {Bergman},
  {Bernath}, {Black}, {Frisk}, {Geppert}, {Hasegawa}, {Hjalmarson}, {Kwok},
  {Larsson}, {Lecacheux}, {Nummelin}, {Olberg}, {Sandqvist}, \&
  {Wirstr{\"o}m}}]{Persson:2007}
{Persson}, C.~M., {Olofsson}, A.~O.~H., {Koning}, N., {et~al.} 2007, \aap, 476,
  807

\bibitem[{{Qi}(2007)}]{Qi:2007}
{Qi}, C. 2007, Advances in Space Research, 40, 639

\bibitem[{{Remijan} {et~al.}(2007){Remijan}, {Markwick-Kemper}, \& {ALMA
  Working Group on Spectral Line Frequencies}}]{Remijan:2007}
{Remijan}, A.~J., {Markwick-Kemper}, A., \& {ALMA Working Group on Spectral
  Line Frequencies}. 2007, BAAS, 38, 963

\bibitem[{{Rodr{\'{\i}}guez} {et~al.}(2005){Rodr{\'{\i}}guez}, {Poveda},
  {Lizano}, \& {Allen}}]{Rodriguez:2005}
{Rodr{\'{\i}}guez}, L.~F., {Poveda}, A., {Lizano}, S., \& {Allen}, C. 2005,
  \apjl, 627, L65

\bibitem[{{Rodriguez Kuiper} {et~al.}(1978){Rodriguez Kuiper}, {Kuiper}, \&
  {Zuckerman}}]{Rodriguez-Kuiper:1978}
{Rodriguez Kuiper}, E.~N., {Kuiper}, T.~B.~H., \& {Zuckerman}, B. 1978, \apjl,
  219, L49

\bibitem[{{Roueff} {et~al.}(2013){Roueff}, {Gerin}, {Lis}, {Wootten},
  {Marcelino}, {Cernicharo}, \& {Tercero}}]{Roueff:2013}
{Roueff}, E., {Gerin}, M., {Lis}, D.~C., {et~al.} 2013, Journal of Physical
  Chemistry A, 117, 9959

\bibitem[{{Schilke} {et~al.}(2001){Schilke}, {Benford}, {Hunter}, {Lis}, \&
  {Phillips}}]{Schilke:2001}
{Schilke}, P., {Benford}, D.~J., {Hunter}, T.~R., {Lis}, D.~C., \& {Phillips},
  T.~G. 2001, \apjs, 132, 281

\bibitem[{{Schilke} {et~al.}(1997){Schilke}, {Groesbeck}, {Blake}, \&
  {Phillips}}]{Schilke:1997}
{Schilke}, P., {Groesbeck}, T.~D., {Blake}, G.~A., \& {Phillips}, T.~G. 1997,
  \apjs, 108, 301

\bibitem[{{Stahl} {et~al.}(2008){Stahl}, {Casassus}, \& {Wilson}}]{Stahl:2008}
{Stahl}, O., {Casassus}, S., \& {Wilson}, T. 2008, \aap, 477, 865

\bibitem[{{Taquet} {et~al.}(2012){Taquet}, {Ceccarelli}, \&
  {Kahane}}]{Taquet:2012a}
{Taquet}, V., {Ceccarelli}, C., \& {Kahane}, C. 2012, \apjl, 748, L3

\bibitem[{{Tercero} {et~al.}(2010){Tercero}, {Cernicharo}, {Pardo}, \&
  {Goicoechea}}]{Tercero:2010}
{Tercero}, B., {Cernicharo}, J., {Pardo}, J.~R., \& {Goicoechea}, J.~R. 2010,
  \aap, 517, A96

\bibitem[{{Tercero} {et~al.}(2011){Tercero}, {Vincent}, {Cernicharo}, {Viti},
  \& {Marcelino}}]{Tercero:2011}
{Tercero}, B., {Vincent}, L., {Cernicharo}, J., {Viti}, S., \& {Marcelino}, N.
  2011, \aap, 528, A26+

\bibitem[{{Turner}(1990)}]{Turner:1990}
{Turner}, B.~E. 1990, \apjl, 362, L29

\bibitem[{{Turner} {et~al.}(1975){Turner}, {Fourikis}, {Morris}, {Palmer}, \&
  {Zuckerman}}]{Turner:1975}
{Turner}, B.~E., {Fourikis}, N., {Morris}, M., {Palmer}, P., \& {Zuckerman}, B.
  1975, \apjl, 198, L125

\bibitem[{{Walmsley} {et~al.}(1987){Walmsley}, {Hermsen}, {Henkel},
  {Mauersberger}, \& {Wilson}}]{Walmsley:1987}
{Walmsley}, C.~M., {Hermsen}, W., {Henkel}, C., {Mauersberger}, R., \&
  {Wilson}, T.~L. 1987, \aap, 172, 311

\bibitem[{{Wang} {et~al.}(2011){Wang}, {Bergin}, {Crockett}, {Goldsmith},
  {Lis}, {Pearson}, {Schilke}, {Bell}, {Comito}, {Blake}, {Caux}, {Ceccarelli},
  {Cernicharo}, {Daniel}, {Dubernet}, {Emprechtinger}, {Encrenaz}, {Gerin},
  {Giesen}, {Goicoechea}, {Gupta}, {Herbst}, {Joblin}, {Johnstone}, {Langer},
  {Latter}, {Lord}, {Maret}, {Martin}, {Melnick}, {Menten}, {Morris},
  {M{\"u}ller}, {Murphy}, {Neufeld}, {Ossenkopf}, {P{\'e}rault}, {Phillips},
  {Plume}, {Qin}, {Schlemmer}, {Stutzki}, {Trappe}, {van der Tak}, {Vastel},
  {Yorke}, {Yu}, \& {Zmuidzinas}}]{Wang:2011}
{Wang}, S., {Bergin}, E.~A., {Crockett}, N.~R., {et~al.} 2011, \aap, 527, A95+

\bibitem[{{Watanabe} \& {Kouchi}(2002)}]{Watanabe:2002}
{Watanabe}, N., \& {Kouchi}, A. 2002, \apjl, 571, L173

\bibitem[{{Watanabe} {et~al.}(2004){Watanabe}, {Nagaoka}, {Shiraki}, \&
  {Kouchi}}]{Watanabe:2004}
{Watanabe}, N., {Nagaoka}, A., {Shiraki}, T., \& {Kouchi}, A. 2004, \apj, 616,
  638

\bibitem[{{Widicus Weaver} \& {Friedel}(2012)}]{Widicus-Weaver:2012}
{Widicus Weaver}, S.~L., \& {Friedel}, D.~N. 2012, \apjs, 201, 16

\bibitem[{{Wootten}(1987)}]{Wootten:1987}
{Wootten}, A. 1987, in IAU Symposium, Vol. 120, Astrochemistry, ed. M.~S.
  {Vardya} \& S.~P. {Tarafdar}, 311--318

\bibitem[{{Zapata} {et~al.}(2011){Zapata}, {Schmid-Burgk}, \&
  {Menten}}]{Zapata:2011}
{Zapata}, L.~A., {Schmid-Burgk}, J., \& {Menten}, K.~M. 2011, \aap, 529, A24

\end{thebibliography}

%
\clearpage

\clearpage

\clearpage

\clearpage

\clearpage

\clearpage
%
%

%
\clearpage

\begin{deluxetable}{llllcc|ccccc|ccccc|}
\rotate
\setlength{\tabcolsep}{0.03in} 
\tablewidth{0pt}
\tabletypesize{\scriptsize}
\tablecolumns{16}
\tablecaption{Spectroscopic and Observational Line Parameters for Formaldehyde\tablenotemark{a,b} \label{tab1}}
\tablehead{ 
\multicolumn{16}{c}{HDCO}\\
\colhead{Frequency} & \colhead{Transition} & \colhead{E$_{up}$} & \colhead{A} &  \multicolumn{2}{c}{Synthesized} &\multicolumn{5}{c}{{dF1}} & \multicolumn{5}{c}{{dF2}} \\
 & &&$\times$10$^{-4}$& \multicolumn{2}{c}{beam}&$\int T_{mb}$dv  & $v_\textnormal{LSR}$ &  $\Delta v_\textnormal{LSR}$ &T$_{ex}$ & N&$\int T_{mb}$dv  & $v_\textnormal{LSR}$ &  $\Delta v_\textnormal{LSR}$&T$_{ex}$ & N \\
(MHz) & & (K) & (s$^{-1}$) & ($\arcsec$ $\times$ $\arcsec$)&PA ($\degr$)& (K~km~s$^{-1}$)& (km~s$^{-1}$) & (km~s$^{-1}$)& (K) & (10$^{14}$cm$^{-2}$)& (K~km~s$^{-1}$)& (km~s$^{-1}$) & (km~s$^{-1}$)& (K) & (10$^{14}$cm$^{-2}$)}
\startdata
192893.26\tablenotemark{c}	& 3$_{0, 3}$--2$_{0, 2}$      &       18.5&  1.94&-- &-- & -- & -- & -- &-- & -- & -- & -- & -- & -- & --\\
193907.46	& 3$_{2, 1}$--2$_{2, 0}$      &       50.4&   1.10&  3.44 $\times$ 3.01&30 & 8.50(1.54)&9.84(0.14)&1.80(0.46) &89(16) &{4.65(0.86)} & 7.38(0.43)&7.75(0.10)& 2.20(0.18)& 56(7) &{2.92(0.22)}\\
193391.61	& 3$_{2, 2}$--2$_{2, 1}$      &       50.4&  1.09&   3.47 $\times$ 3.02 &30&9.05(0.63)&	9.65(0.10)&	1.69(0.16) &89(16)& {4.96(0.37)}&7.07(0.35)&7.57(0.07)&	2.40(0.15)& 56(7) &{2.80(0.18)}\\\cline{7-16}
& & &&& & \multicolumn{5}{c}{{dF3}} & \multicolumn{5}{c}{{dF4}} \\
 & &&&&& $\int T_{mb}$dv  & $v_\textnormal{LSR}$ &  $\Delta v_\textnormal{LSR}$ &T$_{ex}$ & N&$\int T_{mb}$dv  & $v_\textnormal{LSR}$ &  $\Delta v_\textnormal{LSR}$&T$_{ex}$ & N \\ 
 & & &&& & (K~km~s$^{-1}$)& (km~s$^{-1}$) & (km~s$^{-1}$)& (K) & (10$^{14}$cm$^{-2}$)& (K~km~s$^{-1}$)& (km~s$^{-1}$) & (km~s$^{-1}$)& (K) & (10$^{14}$cm$^{-2}$)\\ 
\hline
192893.26\tablenotemark{c}	& 3$_{0, 3}$--2$_{0, 2}$      &       18.5&  1.94& -- & --& -- & -- & -- &-- & -- & -- & -- & -- & -- & --\\
193907.46	& 3$_{2, 1}$--2$_{2, 0}$      &       50.4&   1.10&3.44 $\times$ 3.01&30 &7.13(0.98)&	7.80(0.10)&	1.71(0.32) &57(8) &{2.77(0.42)} & 11.25(1.45)&	8.61(0.05) & 1.13(0.14)& 52(8) &{4.26(0.62)}\\
193391.61	& 3$_{2, 2}$--2$_{2, 1}$      &       50.4&  1.09 &3.47 $\times$ 3.02 &30& 8.00(2.29)&8.00(0.03)&1.39(0.44) &57(8) & {3.12(0.93)}& 9.79(1.02)&	8.30(0.10)&1.66(0.24)& 52(8) &{3.71(0.45)}\\
\hline
\\
\\
\hline
\multicolumn{16}{c}{ (para--)H$_2^{13}$CO}\\
\colhead{Frequency} & \colhead{Transition} & \colhead{E$_{up}$} & \colhead{A} &  \multicolumn{2}{c}{Synthesized} &\multicolumn{5}{c}{{dF1}} & \multicolumn{5}{c}{{dF2}} \\
 & &&$\times$10$^{-4}$& \multicolumn{2}{c}{beam}&$\int T_{mb}$dv  & $v_\textnormal{LSR}$ &  $\Delta v_\textnormal{LSR}$ &T$_{ex}$ & N&$\int T_{mb}$dv  & $v_\textnormal{LSR}$ &  $\Delta v_\textnormal{LSR}$&T$_{ex}$ & N \\
(MHz) & & (K) & (s$^{-1}$) & ($\arcsec$ $\times$ $\arcsec$)&PA ($\degr$)& (K~km~s$^{-1}$)& (km~s$^{-1}$) & (km~s$^{-1}$)& (K) & (10$^{14}$cm$^{-2}$)& (K~km~s$^{-1}$)& (km~s$^{-1}$) & (km~s$^{-1}$)& (K) & (10$^{14}$cm$^{-2}$)\\
\hline
212811.18	& 3$_{0, 3}$--2$_{0, 2}$       &       20.4&    2.64&    3.17 $\times$ 2.75 & 30 &17.99(1.73) & 9.88(0.09)& 1.83(0.24)& 89(16) & { 5.44(0.53)}& 16.04(1.37)&  7.44(0.11) &  2.54(0.29) & 56(7)&{ 2.83(0.26)}\\
213037.34\tablenotemark{d}& 3$_{2, 2}$--2$_{2, 1}$       &       67.7&     1.46  &  -- & --& -- & -- & -- &-- & -- & -- & -- & -- & -- & --\\
213293.57	& 3$_{2, 1}$--2$_{2, 0 }$      &       67.7&   1.46&   3.15 $\times$ 2.75 & 30  & 5.87(0.85)&	9.89(0.19) &	2.21(0.42) &89(16)&{5.43(0.82)}&3.84(0.40) &	7.43(0.13) & 2.16(0.30) & 56(7)& {2.83(0.35)}\\\cline{7-16}
& & &&& & \multicolumn{5}{c}{{dF3}} & \multicolumn{5}{c}{{dF4}} \\
 & & &&& & $\int T_{mb}$dv  & $v_\textnormal{LSR}$ &  $\Delta v_\textnormal{LSR}$ &T$_{ex}$ & N&$\int T_{mb}$dv  & $v_\textnormal{LSR}$ &  $\Delta v_\textnormal{LSR}$&T$_{ex}$ & N \\ 
& & &&& &(K~km~s$^{-1}$)& (km~s$^{-1}$) & (km~s$^{-1}$)& (K) & (10$^{14}$cm$^{-2}$)& (K~km~s$^{-1}$)& (km~s$^{-1}$) & (km~s$^{-1}$)& (K) & (10$^{14}$cm$^{-2}$)\\ 
\hline
212811.18	& 3$_{0, 3}$--2$_{0, 2}$       &       20.4&    2.64&      3.17 $\times$ 2.75 & 30  & 12.93(0.28)&  7.94(0.03) &   1.55(0.04) & 57(8) & { 2.27(0.64)}&9.70(1.16)	&8.44(0.03) & 1.58(0.25) & 52(8)&{ 1.58(0.20)}\\
213037.34\tablenotemark{d}& 3$_{2, 2}$--2$_{2, 1}$       &       67.7&     1.46  & -- & -- & -- & -- & -- &-- & -- & -- & -- & -- & -- & --\\
213293.57	& 3$_{2, 1}$--2$_{2, 0 }$      &       67.7&   1.46&     3.15 $\times$ 2.75 & 30 &  3.14(0.55) &8.03(0.08) &	1.56(0.37) &57(8)& {2.27(0.44)}&2.16(0.26) &	8.63(0.17) &1.30(0.23)& 52(8)&{1.58(0.23)}\\
 \enddata
\tablecomments{The HDCO emission peaks lie at the following positions: dF1 ($\alpha_{J2000}$=05$^{h}$35$^{m}$14$\fs$601, $\delta_{J2000}$=-05$\degr$22$\arcmin$34$\farcs$18), dF2 ($\alpha_{J2000}$=05$^{h}$35$^{m}$14$\fs$054, $\delta_{J2000}$=-05$\degr$22$\arcmin$27$\farcs$82), dF3 ($\alpha_{J2000}$=05$^{h}$35$^{m}$14$\fs$210, $\delta_{J2000}$=-05$\degr$22$\arcmin$42$\farcs$88) and dF4 ($\alpha_{J2000}$=05$^{h}$35$^{m}$13$\fs$861, $\delta_{J2000}$=-05$\degr$22$\arcmin$41$\farcs$92). {The excitation temperature of HDCO towards those peaks is fixed and taken equal to that of H$_2$$^{13}$CO as described in Section~\ref{sec:tempex}.}}
 \tablenotetext{a}{Measured and predicted transitions are available from the CDMS \citep[{http://www.astro.uni-koeln.de/cdms/},][]{Muller:2005} database and at Splatalogue \citep[www.splatalogue.net,][]{Remijan:2007}. More specifically and as previously reported by \citet{Neill:2013a}, we used the spectroscopic data parameters from \citet{Muller:2000} for H$_2^{13}$CO and from \citet{Bocquet:1999} and \citet{Johns:1977} for HDCO.}
\tablenotetext{b}{The numbers in brackets refer to the 1$\sigma$ level uncertainty.} 
\tablenotetext{c,d}{These two lines have been excluded from our analysis since their emission is contaminated by the emission from HCOOCH$_3$ and $^{33}$SO, respectively.} 
     \end{deluxetable}

\clearpage

%
\begin{deluxetable}{rrrrr}
\tablewidth{0.pt}
\tablecolumns{5}
\tablecaption{Total Formaldehyde column densities and D/H ratios\label{tab2}}
\tablehead{ \colhead{Position} & \colhead{N(HDCO)} & \colhead{N(H$_2$$^{13}$CO)}& \colhead{N(H$_2$CO)}&\colhead{D/H ratio}\\
& ($\times$10$^{14}$cm$^{-2}$) & ($\times$10$^{14}$cm$^{-2}$) & ($\times$10$^{16}$cm$^{-2}$) & }
\startdata
dF1 & 4.81(0.62) & 21.70(2.72)& 15.20(1.90) & 0.0032(0.0008)\\
dF2 & 2.86(0.20) & 11.30(1.22)& 7.92(0.86)& 0.0036(0.0006)\\
dF3 & 2.94(0.67) & 9.06(1.00)& 6.34(0.70)& 0.0046(0.0016)\\
dF4& 3.99(0.54) & 6.34(0.85)& 4.44(0.60)& 0.0090(0.0024)\\
 \enddata
 \tablecomments{The total H$_2$$^{13}$CO and H$_2$CO column densities have been derived assuming an ortho:para ratio of 3:1 \citep[e.g. see][]{Kahane:1984,Crockett:2014} and a $^{12}$C/$^{13}$C isotopic ratio of 70 \citep[e.g. see,][]{Favre:2014a}.}
     \end{deluxetable}

\clearpage
%
\begin{deluxetable}{rlrrl}
\tablewidth{0pt}
\tablecolumns{4}
\tablecaption{Spectroscopic parameters of the CH$_3$OH--E methanol transitions observed with the SMA towards the dF3 and dF4 emission peaks\tablenotemark{a} \label{tab3}}
\tablehead{
Frequency & Transition & E$_{up}$ & \colhead{A}   \\
(MHz) & & (K) & ($\times$10$^{-5}$~s$^{-1}$)}
\startdata
193415.37	&	4$_{	0	,	4	}$	--	3$_{	0	,	3	}$	$v_{t}$=0	&	36.3	&	3.03 \\
193441.61	&	4$_{	1	,	4	}$	--	3$_{	1	,	3	}$	$v_{t}$=0	&	28.8	&	2.84 \\
193474.33	&	4$_{	3	,	1	}$	--	3$_{	3	,	0	}$	$v_{t}$=0	&	70.9	&	1.33	\\
193488.99	&	4$_{	3	,	2	}$	--	3$_{	3	,	1	}$	$v_{t}$=0	&	85.9	&	1.33 \\
193506.60	&	4$_{	1	,	3	}$	--	3$_{	1	,	2	}$	$v_{t}$=0	&	44.3	&	2.91	\\
 \enddata
 \tablenotetext{a}{{Measured and predicted transitions are available from the CDMS database \citep{Muller:2005} and at Splatalogue \citep[www.splatalogue.net,][]{Remijan:2007}.}}
      \end{deluxetable}

\end{document}